\documentclass[aps,amsmath,twocolumn,amssymb,floatfix,showpacs,superscriptaddress,nofootinbib,longbibliography]{revtex4-1}
\usepackage{mathtools}
\usepackage{braket}
\usepackage[dvipsnames]{xcolor}
\usepackage{float}
\usepackage{subfigure}
\usepackage{dsfont}
\usepackage[export]{adjustbox}

\usepackage[colorlinks=true,linktoc=page,linkcolor=Purple,citecolor=blue,urlcolor=Violet]{hyperref}

\mathchardef\mhyphen="2D 

\newcommand{\ie}{{i.e.,\,\,}}
\newcommand{\eg}{{e.g.,~}}

\newcommand\bea{\begin{eqnarray}}
\newcommand\eea{\end{eqnarray}}
\newcommand\beq{\begin{equation}}  
\newcommand\eeq{\end{equation}}

\newcommand{\non}{\nonumber}  
\newcommand{\dis}{\displaystyle}

\newcommand{\sg}{\sigma}
\newcommand{\Del}{\Delta}
\newcommand{\Ps}{\Psi}
\newcommand{\UP}{\uparrow}
\newcommand{\DN}{\downarrow}

\newcommand{\mc}{\mathcal}
\newcommand{\mbf}{\mathbf}
\newcommand{\scing}{superconducting }

\usepackage[normalem]{ulem}
\definecolor{lime}{HTML}{A6CE39}
\usepackage{sidecap,tikz}
\DeclareRobustCommand{\orcidicon}{\hspace{-1.0mm}
	\begin{tikzpicture}
		\draw[lime, fill=lime] (0.0,0.0) 
		circle [radius=0.15] 
		node[white] {{\fontfamily{qag}\selectfont \tiny \,ID}};
		\draw[white, fill=white] (-0.0525,0.095) 
		circle [radius=0.007];
	\end{tikzpicture}
	\hspace{-3.0mm}
}
\foreach \x in {A, ..., Z}{\expandafter\xdef\csname orcid\x\endcsname{\noexpand\href{https://orcid.org/\csname orcidauthor\x\endcsname}
		{\noexpand\orcidicon}}
}

\AtBeginDocument{%
	\newwrite\bibnotes
	\def\bibnotesext{Notes.bib}
	\immediate\openout\bibnotes=\jobname\bibnotesext
	\immediate\write\bibnotes{@CONTROL{REVTEX41Control}}
	\immediate\write\bibnotes{@CONTROL{%
			apsrev41Control,author="08",editor="1",pages="1",title="1",year="1"}}
	\if@filesw
	\immediate\write\@auxout{\string\citation{apsrev41Control}}%
	\fi
}%
\begin{document}



\title{Identifying Bogoliubov Fermi surfaces via thermoelectric response in $d$-wave superconductor heterostructure} 

\author{Amartya Pal\orcidA{}}
\email{amartya.pal@iopb.res.in}
\affiliation{Institute of Physics, Sachivalaya Marg, Bhubaneswar-751005, India}
\affiliation{Homi Bhabha National Institute, Training School Complex, Anushakti Nagar, Mumbai-400094, India}

\author{Paramita Dutta\orcidB{}}
\email{paramita@prl.res.in}
\affiliation{Theoretical Physics Division, Physical Research Laboratory, Navrangpura, Ahmedabad-380009, India}

\author{Arijit Saha\orcidC{}}
\email{arijit@iopb.res.in}
\affiliation{Institute of Physics, Sachivalaya Marg, Bhubaneswar-751005, India}
\affiliation{Homi Bhabha National Institute, Training School Complex, Anushakti Nagar, Mumbai-400094, India}


\begin{abstract}
 We theoretically investigate the thermoelectric response of Bogoliubov Fermi surfaces (BFSs) generated in a two dimensional unconventional $d$-wave superconductor subjected to an external in-plane Zeeman field. These BFSs exhibiting the same dimension as the underlying normal state Fermi surface are topologically protected by combinations of discrete symmetries. Utilizing the Blonder-Tinkham-Klapwijk formalism and considering normal-$d$-wave superconductor hybrid junction, we compute the thermoelectric coefficients including thermal conductance, Seebeck coefficient, figure of merit ($zT$), and examine the validation of Widemann-Franz law in the presence of both voltage and temperature bias. Importantly, as a signature of anisotropic nature of $d$-wave pairing, Andreev bound states (ABSs) formed at the normal-superconductor interface play a significant role in the thermoelectric response. In the presence of ABSs, we observe a substantial enhancement in Seebeck coefficient ($\sim 200\,\mu$V/K) and $zT$ ($\sim 3.5$) due to the generation of the BFSs and thus making such setup a potential candidate for device applications. Finally, we strengthen our continuum model results by computing the thermoelectric coefficients based on a lattice-regularized version of our continuum model. 
\end{abstract}

\maketitle

\section{Introduction} \label{Sec:I_intro}

Recent theoretical predictions~\cite{Brydon2016,Timm2017,Agterberg2017,Brydon2018,Yuan2018,Menke2019,Setty2020PRB,Setty2020NatComm} and experimental realizations~\cite{Phan2022,Zhu2021} of Bogoliubov Fermi surfaces (BFSs) in time-reversal symmetry broken (TRSB) superconductors (SCs) and superconducting junctions have spurred interest of the researchers towards this direction~\cite{Link2020b,Link2020a,Oh2020,Oh2021,Timm2021a,Timm2021b,Tamura2020,PhysRevResearch.2.033013,Lapp2020,Herbut2021,
Dutta2021,Bhattacharya2023,Banerjee2022,Pal2024,Miki2024}. These BFSs in superconducting systems, have the same dimensionality as the underlying normal state Fermi surface (FS), which is different from the well-known point or line nodes found in gapless superconductors like $d$-wave superconductors.
The co-existence of both Cooper pairs and Bogoliubov quasiparticles (BQPs) in the superconductors hosting BFSs result in the enhancement of zero-energy single-particle density of states (DoS)~\cite{Agterberg2017,Brydon2018,Yuan2018,Volovik1989JETP} which have been identified as primary signature of BFSs. Importantly, these BFSs are found to be topologically protected by the product of charge conjugation ($\mc{C}$) and parity ($\mc{P}$) symmetry \ie $\mc{CP}$-symmetry and characterized by $\mc{Z}_2$ invariant, providing stability against local perturbations~\cite{Agterberg2017,Brydon2018,Zhao2016}. The combination of elevated zero-energy DoS and topological protection have made these BFSs unique compared to the previously found nodal FSs 
in the literature. 

There are several potential candidate materials to host these BFSs. Initially, BFSs were proposed in multiband SCs hosting interband and intraband pairings or SCs with fermions having effective angular momentum $j=3/2$ like half-Heusler materials~\cite{Brydon2016,Agterberg2017,Brydon2018,Timm2017,Menke2019,Link2020a,Oh2020}. The TRSB pairing or the additional degrees of freedom present in the system induces an effective pseudomagnetic field which is responsible for the  inflation of the point or line nodes in the momentum space~\cite{Brydon2016,Agterberg2017,Brydon2018}. The appearance of any exotic pairing may also have consequence to BFSs~\cite{Dutta2021}. In addition to multiband or spin-$3/2$ systems, BFSs have been theoretically predicted ~\cite{Yuan2018,Setty2020PRB,Setty2020NatComm,Banerjee2022,Cao2023,Pal2024,Wu2024} and experimentally observed~\cite{Zhu2021,Phan2022} in spin-$1/2$ systems consisting of spin-singlet 
$s$-wave SCs. An external TRSB field is required to generate BFSs in spin-$1/2$ systems, in sharp contrast to $j=3/2$ systems where TRSB component is intrinsic in nature. In Ref.\,\cite{Zhu2021}, segmented Fermi surfaces are observed at the surface of a three-dimensional ($3$D) topological insulator placed in close proximity to a $s$-wave SC by applying an external in-plane magnetic field. In Ref.\,\cite{Phan2022}, BFSs are realized in two-dimensional ($2$D) Al-InAs hybrid setup in the presence of external TRSB field. In such heterostructures, BFSs appear when the external field strength is larger than the proximity induced \scing gap but smaller than the gap of parent SC. Although experiments are performed using conventional $s$-wave SC junctions~\cite{Zhu2021,Phan2022}, theoretical models are mostly based on the $d$-wave bulk superconductors~\cite{Setty2020NatComm,Setty2020PRB,Pal2024,Wu2024}. However, heterostructures based on $d$-wave superconductors can also host BFSs when an infinitesimal external magnetic field is applied~\cite{Setty2020PRB,Pal2024}. Signatures of BFSs are  captured in the unconventional $d$-wave SC junction via DoS, differential conductance, and shot-noise spectroscopy~\cite{Setty2020PRB,Pal2024} corroborate with the essential physics obtained using $s$-wave SC junctions~\cite{Yu2018,Banerjee2022}. 

Till date, several proposals for possible experimental detection of these BFSs based on electronic specific heat, thermal conductivity, tunneling conductance, magnetic penetrations depth, NMR spin-lattice relaxation, superfluid density etc. have been reported in the literature employing various models~\cite{Lapp2020,Setty2020PRB,Pal2024,Oh2021}. It is admissible to consider the advantage of the presence of BQPs in addition to Cooper pairs for the identification purposes. Naturally, these BQPs present in those BFSs are expected to carry energy or heat if the system is subjected to the thermal bias. This motivates us to investigate the signatures of BFSs via thermal transport. Importantly, investigation of thermolelectric response in superconducting hybrid structures have recently become a prominent area of research both from the fundamental research and the device application perspectives~\cite{Yokoyama2008,Dutta2017,Paul2016,Jakobsen2020,Kolenda2016,Savander2020,Saxena2023,Dutta2020b,Heidrich2019,Guarcello2023}. On top of that, in a recent work~\cite{Mateos2024}, the nonlocal thermoelectric response in one-dimensional ($1$D) system hosting Bogoliubov-Fermi points have been studied. However, the thermal signatures of BFSs in $2$D systems have not been explored so far, to the best of our knowledge.  Note that, thermal properties of a metal and a $d$-wave SC, in absense of magnetic field, have been addressed in Refs.~\cite{Devyatov2000,Yokoyama2005}, although estimation of Seebeck coefficient and figure of merit are not reported in those articles.

In this article, we intend to fill up this gap and consider a heterostructure comprising of a normal metal and a $d$-wave SC subjected to an in-plane magnetic field hosting BFSs with a $\delta$-function potential barrier at the normal-SC interface tuning the junction transparency. Using the Blonder-Tinkham-Klapwijk (BTK) formalism~\cite{Blonder1982,Riedel1993}, we study the thermoelectric response of BFSs in this hybrid setup after applying a temperature gradient across the system. We compute the charge and thermal current due to applied temperature bias and capture clear signatures of BFSs in them. From the practical perspectives,  we examine Seebeck coefficient and figure of merit by tuning the external magnetic field and junction transparency, and find an enhancement in $zT$ in the transparent limit due to BFSs. We also investigate the validation of Wiedemann-Franz (WF) law, which states that the ratio of the thermal to the electrical conductivity of a great number of metals is directly proportional to the temperature with the proportionality constant, $\mc{L}_0$, known as the Lorenz number~\cite{AshcroftMermin}. Interestingly, $d$-wave SCs host Andreev bound states (ABSs)~\cite{Hu1994,Nagato1995,Tanaka1995,Kashiwaya2000,Tamura2017}, localized at the interface, as a signature of anisotropic nature of $d$-wave pairing. ABSs play a crucial role in electrical transport signatures~\cite{Tanaka1995,Kashiwaya2000,Tamura2017,Pal2024}. In presence of ABSs, we observe a significance enhancement in charge current while minimizing the thermal current, desirable for a better thermoelectric device. Interplay of ABSs and BFSs leads to high Seebeck coefficient ($\sim 200 \mu V/K$) and figure of merit ($\sim 3.5$), suggesting our heterostructure a potential candidate for good thermoelectric device which is one of the main results of the present paper. Finally, we consider a lattice regularized version of our continuum model and perform simulations in the lattice model using the Python package KWANT~\cite{Groth2014} to compute the thermoelectric properties and observe an excellent agreement with the results found using our continuum model.
\begin{figure}
	\includegraphics[width=0.43\textwidth,center]{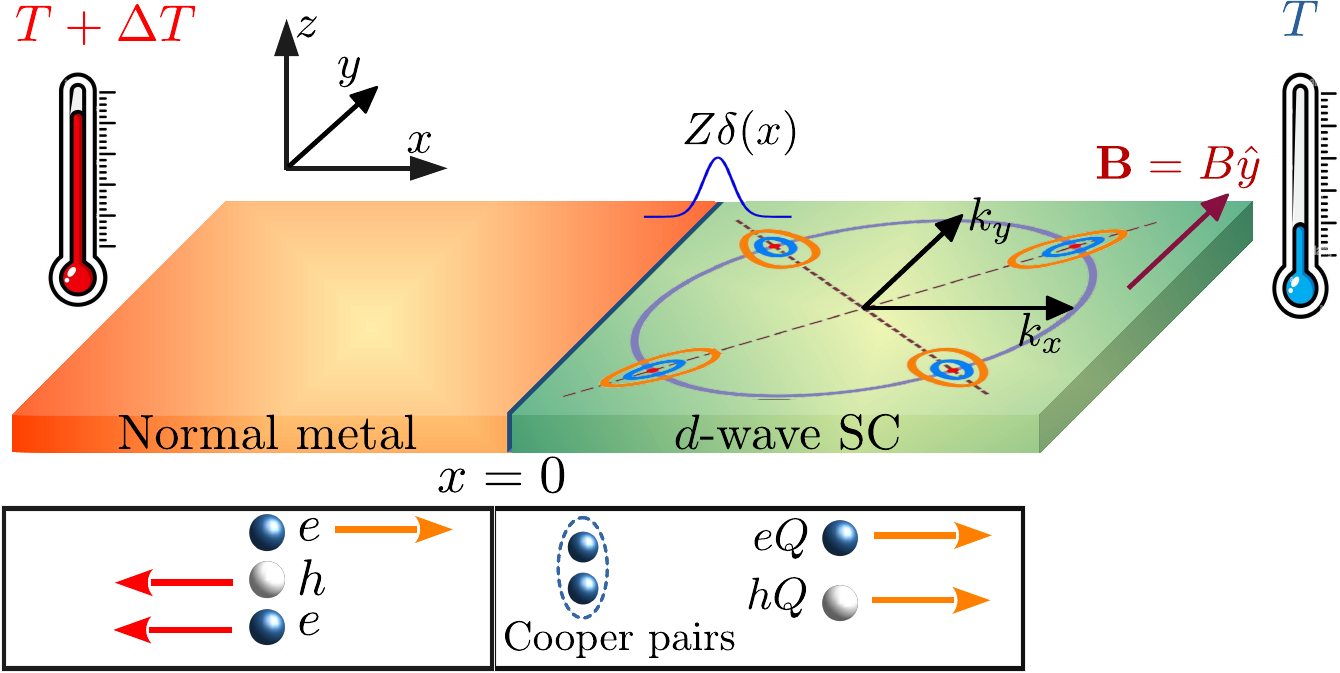}
	\caption{Schematic representation of our hybrid setup consisting of a normal metal ($x<0$) and a $d$-wave SC ($x>0$) with an in-plane magnetic field of strength ($\mbf{B}=B \hat{y}$) 
	is depicted. An interfacial barrier modelled by a $\delta$-function with strength $Z$ (dimensionless) is introduced at the interface ($x=0$). A temperature gradient $\Delta T$ is maintained across 
	the junction. Point nodes originating from $d$-wave order parameter is shown as red dots on top of the SC, whereas the larger grey circle and dotted lines represent the normal metal FS and 
	nodal lines of the bare $d$-wave SC in the absence of the magnetic field respectively. For $B_y>0$, the appearance of BFSs is depicted by the orange elliptic contours on top of the $d$-wave 
	SC. Different scattering processes (at $x=0$) are schemaically shown below the cartoon of our hybrid structure.}
	\label{Fig1.Schematic}
\end{figure}

The remainder of this article is organized as follows. In Sec.~\ref{Sec:II_model}, we introduce the model Hamiltonian of our hybrid setup and outline the theoretical framework. In Sec.~\ref{Sec:III_Results}, we present our numerical results for the thermoelectric coefficients and figure of merit for various model parameters. Finally, we summarize and conclude our paper in Sec.~\ref{Sec:IV_Summary}.


%
\section{Model and Theoretical Framework}\label{Sec:II_model}

We begin with introducing the model under consideration followed by the theoretical framework to compute the thermoelectric properties of our junction.
\subsection{Model Hamiltonian}\label{SubSec:II_modelHam}
We start with a $2$D heterostructure comprised of a normal metal ($x<0$) and a unconventional $d$-wave SC ($x>0$) hosting zero-energy BFSs in the presence of an in-plane magnetic field $\mbf{B}=B\hat{y}$ as schematically shown in Fig.\,\ref{Fig1.Schematic}. The interface at $x=0$ is modelled by a $\delta$-function potential barrier $V(x,y)= V_B \delta(x)$. In reality, this may be possible to implement by an external gate voltage. 

To describe our model, we employ the Bogoliubov-de Gennes (BdG) Hamiltonian using the basis, $\Ps_\mbf{k}=[c_{\mbf{k},\UP},c_{\mbf{k},\DN},c^\dagger_{-\mbf{k},\DN},-c^\dagger_{-\mbf{k},\UP}]^T$, as $H_{{\rm{BdG}}}= \frac{1}{2} \sum_k \Psi_{\mbf{k}}^\dagger\mc{H}(\mbf{k})\Psi_{\mbf{k}} $, where $c_{\mbf{k}\sg}~(c_{\mbf{k}\sg}^\dagger)$ represents the electron annihilation (creation) operator with momentum $\mbf{k}$ ($=\{k_x,k_y\}$) and spin $\sg$. We choose the first quantized Hamiltonian $\mc{H}(\mbf{k})$ as\,\cite{Setty2020PRB,Yang1998,Pal2024},
\begin{equation}
\mc{H}(\mbf{k}) =\xi(\mbf{k}) \tau_z  \sg_0 + \Del(\mbf{k},\alpha) \tau_x \sg_0 -B \tau_0  \sg_y \ , 
\label{Eq:Model_Ham}
\end{equation}
where the Pauli matrices $\mbf{\tau}$ and $\mbf{\sg}$ act on the particle-hole and spin degrees of freedom, respectively. Note that, it is also possible to recast the Hamiltonian in Eq.~\eqref{Eq:Model_Ham} using the BdG basis, $\tilde{\Ps}_\mbf{k}=[c_{\mbf{k},\UP},c_{\mbf{k},\DN},c^\dagger_{-\mbf{k},\UP},c^\dagger_{-\mbf{k},\DN}]^T$ as mentioned in Ref.\,\cite{Sigrist1991}. However, since $\tilde{\Psi}_\mbf{k}$ and $\Psi_\mbf{k}$ are related to each other via an unitary transformation, choice of basis will not affect the results in any way. The kinetic energy of electrons is given by: $\dis{\xi(\mbf{k}) = \frac{\mbf{k}^2}{2m}}-\mu$, resembling a $2$D electron gas. Other terms, $m$ and $\mu$, denote the effective mass of electrons and the chemical potential of the system, respectively. The term $\Del(\mbf{k},\alpha)$ represents $d$-wave \scing pair potential following the explicit form, $ \Del(\mbf{k},\alpha) = \Del_0 \cos[2(\theta + \alpha )]$ with $\theta=\tan^{-1}(k_y/k_x)$ and $\alpha$ denotes the angle between $a$-axis of the crystal and normal to the interface~\cite{Setty2020PRB,Yang1998,Pal2024}.  In presence of the in-plane magnetic field, BFSs appear in the  $x>0$ region as depicted in Fig.~\ref{Fig1.Schematic}. To find the thermoelectric properties of BFSs, we add the normal metal region to form the metal-SC junction. For the normal metal region, we consider the same kinetic energy $\xi(\mbf{k})$ considering $B=0$ and $\Delta ({\bf k},\alpha)=0$. 

Note that, these FSs are protected by the combined change conjugation and parity ($\mc{CP}$) symmetry which make them stable against all $\mc{CP}$ preserving perturbations~\cite{Agterberg2017,Bhattacharya2023}. Moreover, the topological protection of these FSs are also identified by computing Pfaffian, $Pf(k)$, which exhibits opposite sign inside and outside the FSs. 
For this model, $Pf(\mbf{k})=\xi(\mbf{k})^2 + \Del(\mbf{k})^2 - B_y^2$ \cite{Pal2024,Setty2020PRB}. Using Pfaffian, a $\mc{Z}_2$ invariant, $\nu$, defined as $(-1)^\nu=sgn[Pf(\mbf{k})Pf(0)]$, can also be assigned to characterize the topological nature of these FSs \cite{Agterberg2017,Bhattacharya2023,Pal2024}. BFSs seperate the $\nu=0$ and $\nu=1$ regions in the momentum space. However, in this article, our main objective is to study the thermoelectric signature of BFSs in the heterojunction as presented in Fig.~\ref{Fig1.Schematic}. The band structure, Fermi surface, and density of states (DoS) for each of the four bands considering various crystal orientation angle ($\alpha$) and magnetic field are discussed in detail in Ref.\,\cite{Pal2024}.

Throughout our study, we fix $m=\Delta_0=\hbar=1$, and set $\mu=10\Del_0$, $\alpha=0$, and $\alpha=\pi/4$. Specifically, the $d$-wave pair potential takes the form, $\Delta({\mbf{k},\alpha=0})=\Delta_0 (k_x^2 -k_y^2)/k_f^2$ and $\Delta({\mbf{k},\alpha=\pi/4})=2\Delta_0 k_x k_y/k_f^2$, following the symmetry of $d_{x^2-y^2}$ and $d_{xy}$ orbital, respectively, with the Fermi momentum 
$k_f=\sqrt{2m\mu}$~\cite{Hu1994,Tanaka2021,Zhu1999}. The pair potential is temperature-dependent following the relation $\Delta(\mbf{k},T)=\Delta(\mbf{k}) \tanh (1.74\sqrt{T_c/T-1})$~\cite{Enoksen2012,Tao2012,Ren2013,Huang2023}, where $T_c$ is the critical temperature of the superconductor.

\subsection{Theoretical framework} \label{SubsecII_formalism}
Within the linear response regime, the electrical and thermal current in presence of small voltage difference $\Delta V$ and temperature gradient $\Delta T$ can be expressed as~\cite{Riedel1993}, 
\begin{equation}
\begin{bmatrix}
	~ I_q \\I_{th} 
\end{bmatrix} = \begin{bmatrix}
~ L_{11}  &L_{12} \\
~ L_{21} & L_{22} 
\end{bmatrix}  \begin{bmatrix}
	~ \Delta V \\ \Delta T 
\end{bmatrix}\ ,
\end{equation}
where, $L_{11}$ and $L_{22}$ denote the electrical and thermal conductance describing the charge and heat current carried by the system in application of the applied voltage bias and temperature gradient, respectively. The off-diagonal element of the matrix \ie $L_{12}$ ($L_{21}$) is the thermoelectric coefficient representing the charge (heat) current flow caused by the temperature gradient (voltage bias). These coefficients are expressed as~\cite{Riedel1993,Sivan1986,Saxena2023,Wysokinski_2012,Dutta2017,Dutta2020b,Dutta2023,Saxena2023},
\begin{eqnarray}
L_{11}&=& \frac{e^2}{h} \int_{0}^{\infty} dE \,\mc{T}_e(E) \left(-\frac{\partial f (E,T)}{\partial E}\right) \label{Eqn. L11},	\\
L_{12}&=& \frac{e}{hT}\int_{0}^{\infty} dE \,\mc{T}_e(E) E \left(-\frac{\partial f (E,T)}{\partial E}\right),  \label{Eqn. L12} \\
L_{22}&=& \frac{1}{hT} \int_{0}^{\infty} dE \,\mc{T}_{th}(E) E^2 \left(-\frac{\partial f (E,T)}{\partial E}\right)  ,
\label{Eqn. L22}
\end{eqnarray}
where $\mc{T}_e(E)$ and $\mc{T}_{th}(E)$ denote transmission functions for charge and thermal current flow, respectively and given by,
\begin{eqnarray}
 \mc{T}_e(E) \!=\!\sum_{\sg=\UP,\DN}\int_{-\pi/2}^{\pi/2}d\theta_e \cos\theta_e [1-\mc{R}^e_\sg(\theta_e) + \mc{R}^h_{\sg}(\theta_e)]\ , \non  \\ 
   \label{Eqn.T_e(E)}\\
  \mc{T}_{th}(E) \!=\!\sum_{\sg=\UP,\DN}\int_{-\pi/2}^{\pi/2}d\theta_e \cos\theta_e [1-\mc{R}^e_\sg(\theta_e) - \mc{R}^h_{\sg}(\theta_e)]\ . \non \\ \label{Eqn.T_th(E)}
\end{eqnarray}
Here, $\mc{R}^e_\sg\! (\theta_e)=\sum_{\bar{\sg}=\UP,\DN} |r^{ee}_{\bar{\sg},\sg}(\theta_e)|^2 $ and  $\mc{R}^h_{\sg}(\theta_e)=\sum_{\bar{\sg}=\UP,\DN}\frac{\cos\theta_h}{\cos\theta_e}|r^{eh}_{\bar{\sg},\sg}(\theta_e)|^2$ correspond to the probability of normal reflection (NR) and Andreev reflection (AR) respectively, for an incident electron with spin $\sg$ and incident angle $\theta_e$. 
In Eqs.\,\eqref{Eqn. L11}-\eqref{Eqn. L22}, $L_{11}$, $L_{12}$, and $L_{22}$ are expressed in units of $e^2/h$, $k_Be/h$, and $k_B^2T/h$, respectively, where, $e$, $k_B$ and $h$ are the electron charge, Boltzmann constant and Planck's constant, respectively. Here, $f(E,T)$ is the Fermi-Dirac distribution function of the thermal reservoir at temperature $T$. In Sec.\,\ref{Sec:III_Results}, we depict the behaviour of $L_{12}$ and $L_{22}$ after normalizing them by $L_{12}^0$ and $L_{22}^0$ respectively. The normalization constants, $L_{12}^0$ and $L_{22}^0$ are obtained using the Eq.~\eqref{Eqn. L12} and Eq.~\eqref{Eqn. L22}, after replacing SC ($x>0$ in Fig.~\ref{Fig1.Schematic}) by a normal metal with zero magnetic field \ie setting $\Del_0=0$ and $B_y=0$. Since, SC is absent for $L_{12}^0$ and $L_{22}^0$, probability of AR ($\mc{R}_{\sg}^h$) is also zero in Eqs.~\eqref{Eqn.T_e(E)} and $\eqref{Eqn.T_th(E)}$.

For the application point of view, we also study the Seebeck coefficient $\mc{S}$ which measures the open circuit voltage ($\Delta V$) developed across the junction due to the applied temperature gradient ($\Delta T$), and figure of merit $zT$ defined as, \cite{Wysokinski_2012,Dutta2017,Saxena2023},
\begin{eqnarray}
	\mc{S}&=&-\frac{\Delta V}{\Delta T} = \frac{L_{12}}{L_{11}}\ , \label{Eq. Seebeck}\\
	zT&=& \frac{\mc{S}^2L_{11}T}{L_{22}-L_{12}^2/(L_{11}T)} \label{Eq. zT}\ .
\end{eqnarray} 
where $\mc{S}$ is expressed in units of $k_B/e$ and $zT$ is a dimensionless quantity characterizing the efficiency of the thermoelectric device.

The behavior of the thermoelectric response is primarily governed by transmission functions, $\mc{T}_e$ and $\mc{T}_{th}$, for a particular $T/T_c$ as the model parameters exhibit direct influence 
on these functions via different scattering probabilities. Furthermore, $\mc{T}_e$ and $\mc{T}_{th}$ have two major components in them, probability of NR ($\mc{R}_{\sg}^e$) and  AR ($\mc{R}_{\sg}^h$). Thus, the interplay of NR and AR regulates all the responses under thermal bias. Additionally, the conservation of probability current demands the following unitarity relation:
\begin{equation}
\mc{R}_{\sg}^e + \mc{R}_{\sg}^h + \mc{T}_{\sg}^e =1 \label{Eq.Unitarity}
\end{equation}
 where $\mc{T}_{\sg}^e=\sum_{\bar{\sg}, \gamma} \!|t_{\bar{\sg},\sg}^{\gamma e}|^2$ with $\bar{\sg}=\{\UP,\DN\}$, $\gamma=\{e,h$\} and $\mc{T}_{\sg}^e$ is the probability of an electron with spin $\sg$ in the normal metal to transmit inside the superconductor. We also note that, in Eq.~\eqref{Eqn.T_e(E)} and Eq.~\eqref{Eqn.T_th(E)}, $\mc{R}_{\sg}^h$ appears with opposite signs. This is because the electrical conductance is sensitive to the sign of charged particles whereas thermal conductance is not affected by the sign of electric charge as that is an energy current. We compute all these 
 scattering coefficients using scattering matrix formalism. We refer to Appendix\,\ref{Appendix_scattering_Matrix} for the details of our scattering matrix calculation.
 
\section{Results and Discussion}\label{Sec:III_Results}
In this section, we present and discuss our numerical results for $L_{12}, L_{22}, \mc{S}$, and $zT$, and examine the validation of WF law in the superconducting heterostructure under consideration 
for $\alpha=0$ and $\pi/4$. Finally, we recheck and support the behaviors of these thermoelectric coefficients using a tight binding model obtained by discretizing our continuum model.

\subsection{$\alpha=0$ : $d_{x^2-y^2}$ pairing}\label{Subsec:III_a}
In this subsection, we present our numerical results considering $d_{x^2-y^2}$ pairing ($\alpha=0$) of the $d$-wave SC involved in the junction.
\subsubsection{Thermoelectric coefficient and thermal conductance}

\begin{figure}
	\includegraphics[scale=0.49]{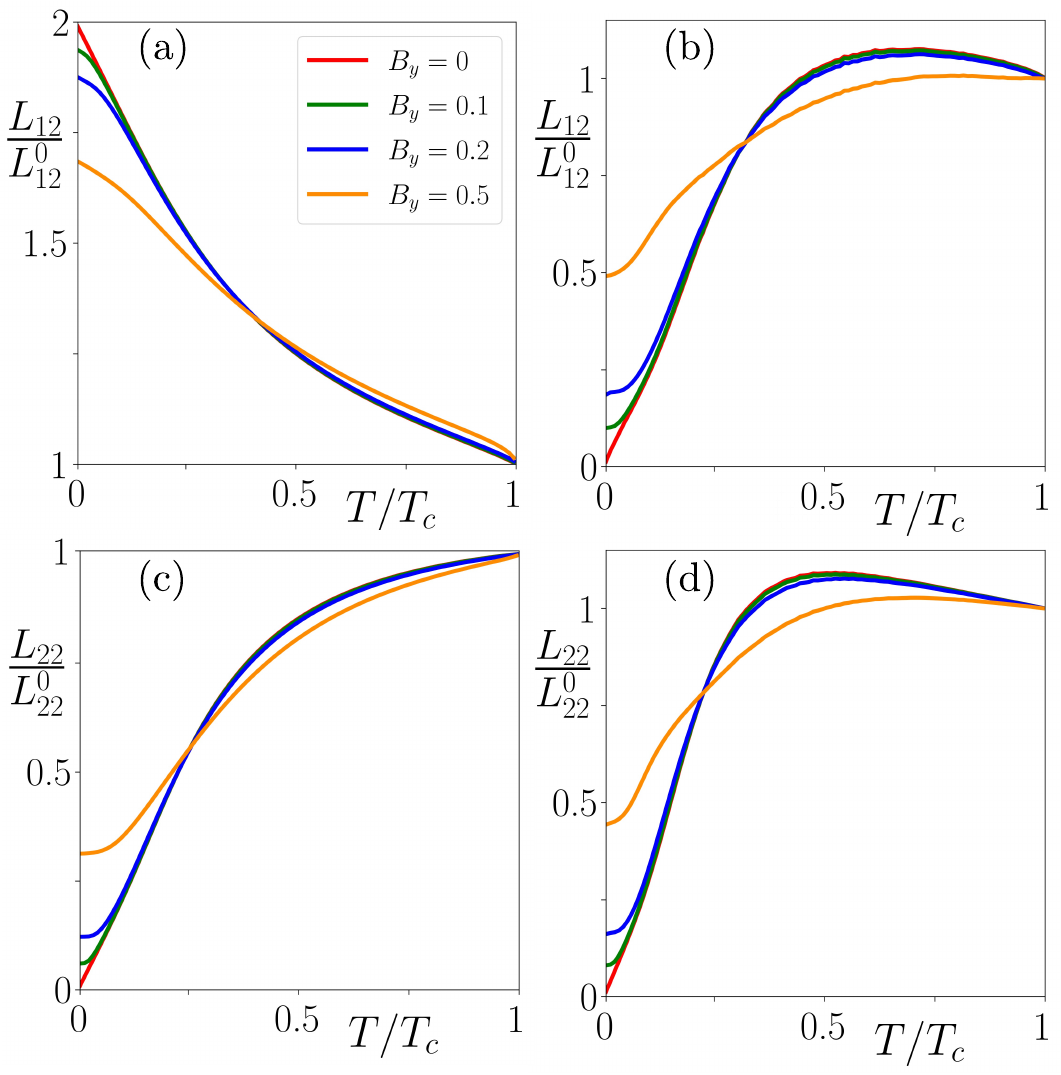}
	\caption{In panels (a-b), normalized thermoelectric coefficient $L_{12}/L_{12}^0$ and in panels (c-d), thermal conductance $L_{22}/L_{22}^0$ are shown respectively as a function of temperature choosing different values of the magnetic field. Here, panels (a,c) and panels (b,d) correspond to the transparent limit ($Z=0$) and tunneling limit ($Z=10$), respectively for $\alpha=0$.}
	\label{Fig2.L12_L22_vs_temp}
\end{figure}

We depict the behaviour of normalized $L_{12}$ and $L_{22}$ as a function of temperature normalized by $T_c$ \ie $T/T_c$, in Fig.\,\ref{Fig2.L12_L22_vs_temp}(a-b) and Fig.\,\ref{Fig2.L12_L22_vs_temp}(c-d), respectively, for various strengths of the in-plane magnetic field (measured in units of $\Delta_0$). As mentioned before, we normalize $L_{12}$ and $L_{22}$ by the normalization constant, $L_{12}^0$ and $L_{22}^0$ respectively. 
In Fig.\,\ref{Fig2.L12_L22_vs_temp}(a-b), the normalized $L_{12}$ is depicted for the two limits of the junction transparency \ie transparent limit ($Z=0$) and tunneling limit ($Z=10$), respectively. In Fig.\,\ref{Fig2.L12_L22_vs_temp}(a), for $Z=0$ and $B_y=0$, we observe that the thermoelectric coefficient takes the value $2$ at $T=0$ and gradually decreases with the increase of the temperature. Note that, the probability of NR is zero ($\mc{R}_{\sg}^e=0$) in the transparent limit. Thus, using the unitarity relation [see Eq.\eqref{Eq.Unitarity}] one can recast 
Eq.~\eqref{Eqn.T_e(E)} as,
\begin{eqnarray}
\mc{T}_e(E) &=& \sum_{\sg}\int_{-\pi/2}^{\pi/2}d\theta_e \cos\theta_e [1 +\mc{R}_{\sg}^h(E)]\ , \non \\ &=&\sum_{\sg}\int_{-\pi/2}^{\pi/2}d\theta_e \cos\theta_e [2- \mc{T}_{\sg}^e(E)]\ .
  \label{Eq.T_q_Z0_dx2y2}
 \end{eqnarray}
Thus when AR $\mc{R}_{\sg}^h=1$, the thermoelectric coefficient becomes $2$ at $T=0$. However, it saturates to the unity when the temperature is close to the critical temperature where the superconducting gap vanishes completely and the behavior of the SC reverts back to the normal metal phase. As soon as we apply the magnetic field, BFSs are generated in the SC enhancing $\mc{T}_{\sg}^e(E)$ due to the increased number of BQPs within the subgapped regime. Thus, $L_{12}/L_{12}^0$ decreases with the magnetic field. However, increasing $B_y$ leads to the reduction of normalized $L_{12}$ up to a certain crossover temperature $T_r$ ($\sim 0.4$ in this case) above which the opposite behaviour is observed. This behavior is due to the fact that for $T>T_r$, QP states above $E>\Delta_0$ starts contributing significantly to the thermoelectric responses dominating the contribution of BFSs. We confirm this by calculating the individual contributions arising from the sub-gap and above-gap states (see Appendix~\ref{Append_A} for details). 

In the tunneling limit ($Z=10$), the scenario [see Fig.\,\ref{Fig2.L12_L22_vs_temp}(b)] becomes exactly opposite. In this limit, the probability of AR is strongly suppressed by NR due to the presence 
of the strong barrier potential at the junction. Thus, the transmission function becomes,
\begin{eqnarray}
 \mc{T}_e(E) &\simeq& \sum_{\sg}\int_{-\pi/2}^{\pi/2}d\theta_e \cos\theta_e [1 - \mc{R}_{\sg}^e(E)]\ ,\non \\
  &\simeq&  \sum_{\sg}\int_{-\pi/2}^{\pi/2}d\theta_e \cos\theta_e \,\mc{T}_{\sg}^e(E)\ .
  \label{Eq.T_q_Z10_dx2y2}
\end{eqnarray} 
Let us now focus on the low temperature limit i.e., close to $T/T_c=0$. In the absence of the magnetic field, $L_{12}$ becomes vanishingly small following $R_{\sg}^e\simeq 1$ or $T_{\sg}^e\simeq0$. 
In the presence of the magnetic field, the generation of BFSs in the system leads to the finite value of $\mc{T}_{\sg}^e$ which increases with enhancing the strength of the field. With the 
rise of temperature, $L_{12}/L_{12}^0$ increases till the crossover temperature in sharp contrast to the ballistic case. Note that, $\mc{T}_{\sg}^e(E)$ appears with opposite sign in the expression of $\mc{T}_e(E)$ in the ballistic and tunneling limit, due to which such opposite behaviour is observed [see Eqs.~\eqref{Eq.T_q_Z0_dx2y2} and \eqref{Eq.T_q_Z10_dx2y2}]. These observations are central to this paper and are not reported earlier in the literature.

In Fig.\,\ref{Fig2.L12_L22_vs_temp}(c-d), we observe that the thermal conductance increases with the rise of temperature irrespective of the barrier strength. This variation can be attributed to the fact that as we increase the temperature, the occupation of QPs with the higher energy becomes more probable following the Fermi-Dirac distribution function $f(E,T)$ and their contributions to the thermal conductance also increase. Note that, our results corroborate with the findings reported in Refs.~\cite{Devyatov2000,Yokoyama2005} for $B_y=0$. In the presence of finite magnetic field, we note that the thermal conductance increases until the crossover temperature $T_r$ ($\simeq 0.2$ in this case) is achieved, but exhibit the opposite behavior when it is above the crossover regime \ie $T>T_r$. We refer to Appendix\,\ref{Append_A} for the detailed analysis of the contributions arising from different energy window. The enhancement of $L_{22}/L_{22}^0$ at a fixed $T$ (below $T<T_r$) can be explained using the transmission function for the thermal current $\mc{T}_{th}(E)$. 
Mathematically, in the expression of $\mc{T}_{th}$ [Eq.\eqref{Eqn.T_th(E)}], both $R_{\sg}^e$ and $R_{\sg}^h$ appear with the same sign. Thus, utilizing the unitarity relation, we can write,
 \begin{eqnarray}
 \mc{T}_{th}(E) &=& \sum_{\sg}\int_{-\pi/2}^{\pi/2}d\theta_e \cos\theta_e [1 - \mc{R}_{\sg}^e(E) - \mc{R}_{\sg}^h(E)]\ , \non \\ &=& \sum_{\sg}\int_{-\pi/2}^{\pi/2}d\theta_e \cos\theta_e \,\mc{T}_{\sg}^e(E)\ .
  \end{eqnarray}
 in case of both the transparent and tunneling limit. As stated earlier, finite value of $B_y$ leads to the generation of BFSs which enhances $\mc{T}_{\sg}^e(E)$. Thus, the enhancement of $L_{22}/L_{22}^0$ takes place as one increases the strength of $B_{y}$ irrespective of the strength of the barrier potential. Such enhancement (below $T<T_r$) of thermal conductance is due to the BFSs as shown in both Fig.\,\ref{Fig2.L12_L22_vs_temp}(c) and Fig.\,\ref{Fig2.L12_L22_vs_temp}(d). Interestingly, in Fig.\,\ref{Fig2.L12_L22_vs_temp}(b),(c), and (d), for $B_y=0$, $L_{12}$ and $L_{22}$ vanishes at $T=0$. The reason can be attributed to the following. The transmission functions in Eq.~\eqref{Eqn.T_e(E)} and \eqref{Eqn.T_th(E)} are related to the transission probability 
$\mc{T}_{\sg}^e(E)$ which is proportional to the DoS inside the SC gap. For $B_y=0$, zero-energy DoS inside the SC is vanishingly small~\cite{Pal2024}. Furthermore, at $T=0$, system can only acess the states at the Fermi surface since thermal excitations are absent. This is why for $B_y=0$, $L_{12}$ and $L_{22}$ in Fig.\,\ref{Fig2.L12_L22_vs_temp}(b),(c), and (d) vanish at $T=0$.

\begin{figure}
	\includegraphics[scale=0.49]{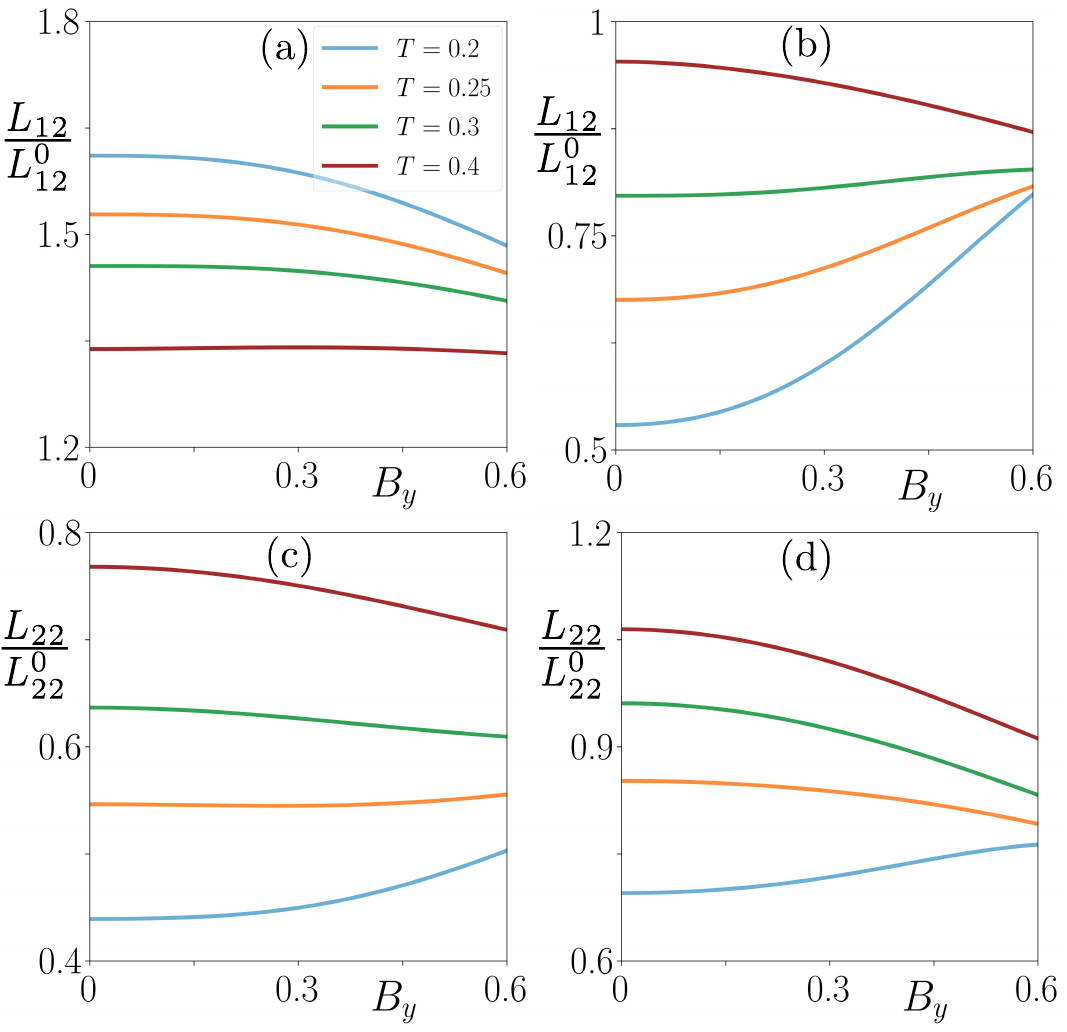}
	\caption{In panels (a-b), normalized thermal coefficient and in panels (c-d), thermal conductance are depicted as a function of the magnetic field $B_y$ respectively, for different values of temperatures 
	$T/T_c$. Here, panels (a,c) and panels (b,d) correspond to ballistic ($Z=0$) and tunneling ($Z=10$) limit respectively, with $\alpha=0$.}
	\label{Fig3.L12_L22_vs_By}
\end{figure}
In order to understand the behaviour of these thermoelectric coefficients with the magnetic field in more detail, we now discuss those quantities as a function of $B_y$ choosing various values of 
$T/T_c$ and depict them in Fig.\,\ref{Fig3.L12_L22_vs_By}. For sufficiently low temperatures ($T<T_r$), the states with energy $E<\Delta_0$ significantly contribute to the thermal current (see Appendix\,\ref{Append_A} for details). Therefore, to investigate the signature of BFSs, being the zero-energy excitations, we focus on temperatures less than the crossover temperature. For $T/T_c=0.2$ [see Fig.\,\ref{Fig3.L12_L22_vs_By}(a)], we observe that $L_{12}/L_{12}^0$ reduces with increasing magnetic field whereas in Fig.\,\ref{Fig3.L12_L22_vs_By}(b), we notice the opposite behavior \ie the enhancement of the same with increasing magnetic field. The reason can be attributed to the generation of BFSs in the system, leading to the elevation in the transmission probability $T_{\sg}^e$ which reduces $\mc{T}_e(E)$ for $Z=0$ and enhances $\mc{T}_e(E)$ in case of $Z=10$. This behaviour is persistent up to $T<T_r$ ($\sim 0.4$ in this case), above which QPs with $E>\Delta_0$ substantially contribute screening the signatures of BFSs.

In Fig.\,\ref{Fig3.L12_L22_vs_By}(c) and \ref{Fig3.L12_L22_vs_By}(d), the normalized $L_{22}$ is shown as a function of $B_y$ for $Z=0$ and $10$, respectively. For both the barrier strength $Z$, 
we observe the enhancement of $L_{22}$ only for $T/T_c=0.2$. Above the crossover temperature, $T_r$ ($\sim 0.25$ in this case), $L_{22}$ reduces with the increase in the magnetic field. The enhancement of $L_{22}$ originates from the generation of BFSs which further increases the thermal transmission function $\mc{T}_{th}(E)$ with the enhancement of the magnetic field strength.

\subsubsection{Seebeck coefficient, figure of merit, and WF law}

\begin{figure}
	\includegraphics[width=0.5\textwidth,left]{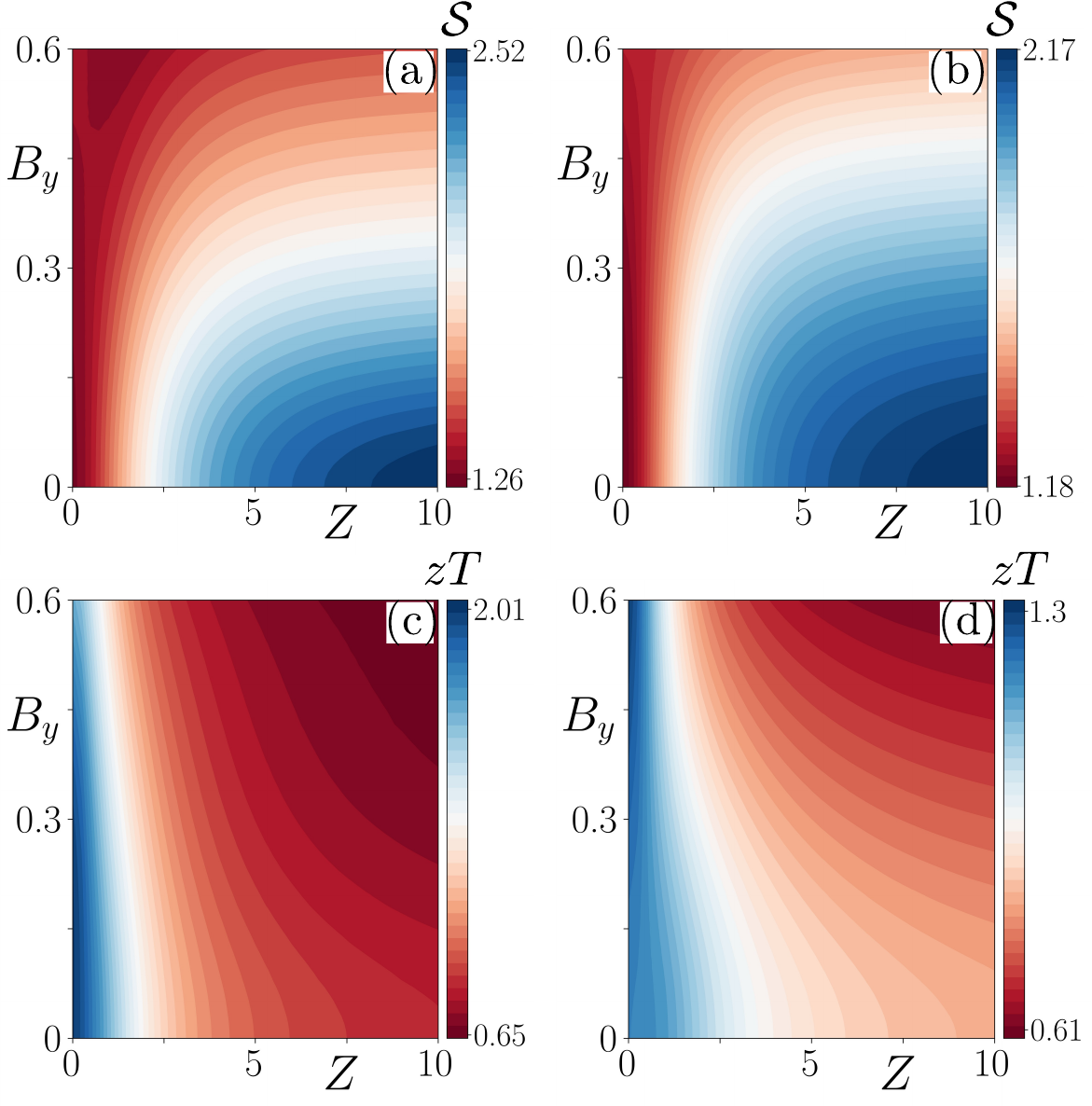}
	\caption{(a-b) Seebeck coefficient (measured in units of $k_B/e$) and (c-d) figure of merit, $zT$, are demonstrated in the $B_y\!-Z$ plane choosing $T/T_c=0.2$ (a,c) and $T/T_c=0.4$ (b,d) 
	considering $\alpha=0$.}
	\label{Fig4_Seebeck_zT_alpha0}. 
\end{figure}

With the understanding of the features of the thermal coefficient and the thermal conductance, we now turn our attention to the physical quantities which carry significance from the perspective of the device applications. We compute Seebeck coefficient ($\mc{S}$) using Eq.~\eqref{Eq. Seebeck} and figure of merit ($zT$) using Eq.~\eqref{Eq. zT}, and depict their behavior in the $B_y-Z$ plane in Fig.\,\ref{Fig4_Seebeck_zT_alpha0}. As we intend to investigate the signature of BFSs,  we fix the temperature at lower values ($T<T_{r}$). We observe that Seebeck coefficient decreases with the increase in $B_y$ irrespective of the value of $Z$. However, it increases with the rise of the barrier strength for a particular value of magnetic field. We find that Seebeck coefficient attains a maximum value $\sim 2.5 k_B/e$ for $T/T_c=0.2$ [see Fig.~\ref{Fig4_Seebeck_zT_alpha0}(a)] and $\sim 2.2 k_B/e$ for $T/T_c=0.4$ as shown in 
Fig.~\ref{Fig4_Seebeck_zT_alpha0}(b). The behavior of the Seebeck coefficient is decided by the ratio of the thermoelectric coefficient to the electrical conductance. Although large values of $S$ implies large conversion rate of heat energy into electrical current, nevertheless it does not incorporate in the thermal conductance, $L_{22}$ [see Eq.~\eqref{Eq. Seebeck}]. Devices with large $L_{22}$ is not desirable for thermoelectric applications as it gives rise to Joule heating and reduces performance. Thus, to estimate the efficiency of the thermoelectric device more accurately, $zT$ becomes the suitable quantity to compute. We showcase $zT$ as a function of $B_y$ and $Z$ by fixing $T/T_c$. We observe $zT$ takes maximum value $\sim 2$ for $T/T_c=0.2$ and $\sim 1.3$ for $T/T_c=0.4$. Interestingly, for $T/T_c=0.4$, the enhancement of $zT$ is noticed in the ballistic limit by increasing the strength of $B_y$ \ie generation of BFSs reduce $L_{22}$ and thus help to improve the device efficiency. Also, in the tunneling limit $\mc{S}$ becomes larger for small $B_{y}$ [blue region in Fig.~\ref{Fig4_Seebeck_zT_alpha0}(b)] giving rise to relatively higher value of $zT$ compared to $T/T_c=0.2$. Note that, $S$ is always positive due to contributions by electrons only.

\begin{figure}
	\includegraphics[scale=0.49]{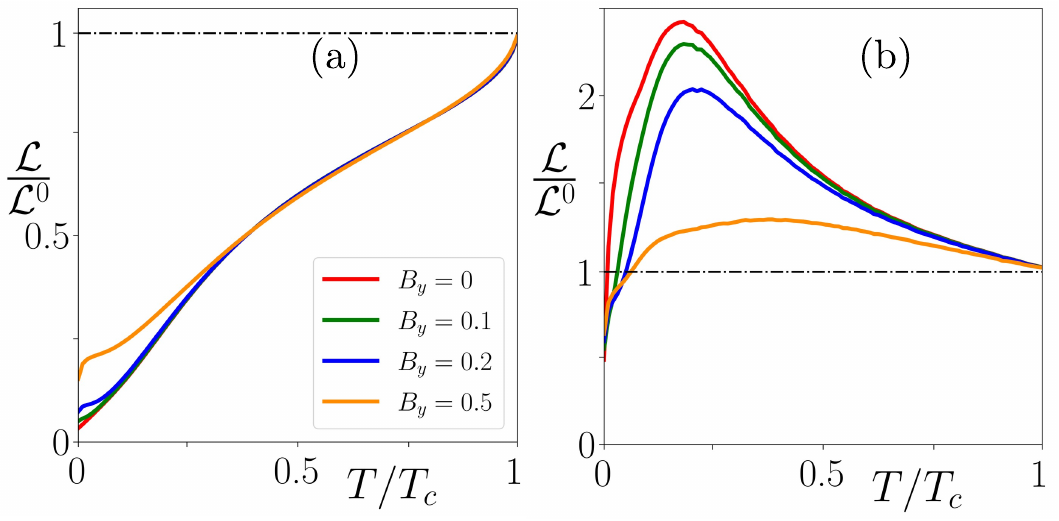}
	\caption{Variation of the normalized Lorenz number ($\mc{L}/\mc{L}_0$) is shown as a function of temperature ($T/T_c$) considering various magnetic fields ($B_y$) in panel (a) ballistic $(Z=0)$ 
	and panel (b) tunneling limit $(Z=10)$. Here, $\mc{L}_0=\frac{\pi^2}{3}(k_B/e)^2$ denotes the Lorentz number for ideal Fermi gas.}
	\label{Fig5.Lorenz_number}
\end{figure}

We also examine the validation of WF law in our superconducting heterostructure. Violation of WF law is reported earlier in the literature~\cite{Wakeham2011,Tanatar2007,Crossno2016,Ghanbari2013,Yokoyama2005} for other systems. To investigate the violation of WF law, we numerically compute the Lorenz number $\mc{L}$, calculated from the ratio of $L_{11}$ to $L_{22}$, and normalize it by the Lorenz number for the free Fermi gas, $\mc{L}_0=\pi^2/3 (k_B/e)^2$ \cite{Franz1853}. We present the numerical results for $\mc{L}/\mc{L}_0$ as a function of $T/T_c$ for various strengths of $B_y$ in the ballistic and tunneling limit in Fig.\,\ref{Fig5.Lorenz_number}(a-b). For $T/T_c$ close to zero, the deviation of $\mc{L}$ from $\mc{L}_0$ is maximum and this deviation reduces as $T/T_c$ approaches to $1$. This is because when 
the temperature is close to $T/Tc=1$, the superconducting gap almost vanishes and the SC behaves as a normal metal. Thus, as $T/T_c\rightarrow 1$, the Lorenz number becomes equal to the Lorenz number of the free Fermi gas (metal). Note that, in presence of the magnetic field, the deviation of $\mc{L}$ from $\mc{L}_0$ reduces for the intermediate values of the temperature in both the limits. Interestingly, $\mc{L}/\mc{L}_0$ is less than 1 for $Z=0$ while it is mostly greater than 1 for $Z=10$. The value of $\mc{L}/\mc{L}_0$ greater than 1 indicates the dominance of $L_{11}$ ($\sim L_{12}$ roughly) over $L_{22}$ which is desirable for good thermoelectric device. This feature is also reflected in $\mc{S}$ as shown in Fig.~\ref{Fig4_Seebeck_zT_alpha0}(a-b) where $S$ becomes large in the tunneling limit.

\subsection{$\alpha=\pi/4$: $d_{xy}$ pairing}\label{Subsec:III_b}
In this subsection, we present our results for $L_{12}$, $L_{22}$, $\mc{S}$, and $zT$ considering $d_{xy}$ pairing ($\alpha=\pi/4$) of $d$-wave SC concerning our heterostructure. The additional 
feature of $d_{xy}$ pairing, compared to $d_{x^2-y^2}$ pairing, is the formation of zero energy interfacial localized ABSs in the tunneling limit, as a signature of anisotropic pairing of $d$-wave SC \cite{Hu1994,Nagato1995,Tanaka1995}. Importantly, the formation of such ABSs leads to the appearance of zero energy conductance peak (ZBCP) in the differential conductance profile~\cite{Tanaka1995,Kashiwaya2000,Tamura2017,Yokoyama2005,Pal2024} originating from multiple Andreev reflections at the interface~\cite{Kashiwaya2000}. Interestingly, the presence of an external magnetic field can shift the location of the conductance peak from zero to the finite energy~\cite{Pal2024} which helps identifying the signature of BFSs present at zero energy. Motivated by this, we explore the combined effect of the ABSs and BFSs on the thermoelectric response in our concerned $d$-wave SC hybrid junction set up.

\subsubsection{Thermoelectric coefficient and thermal conductance}
%
\begin{figure}[]
	\includegraphics[width=0.49\textwidth,left]{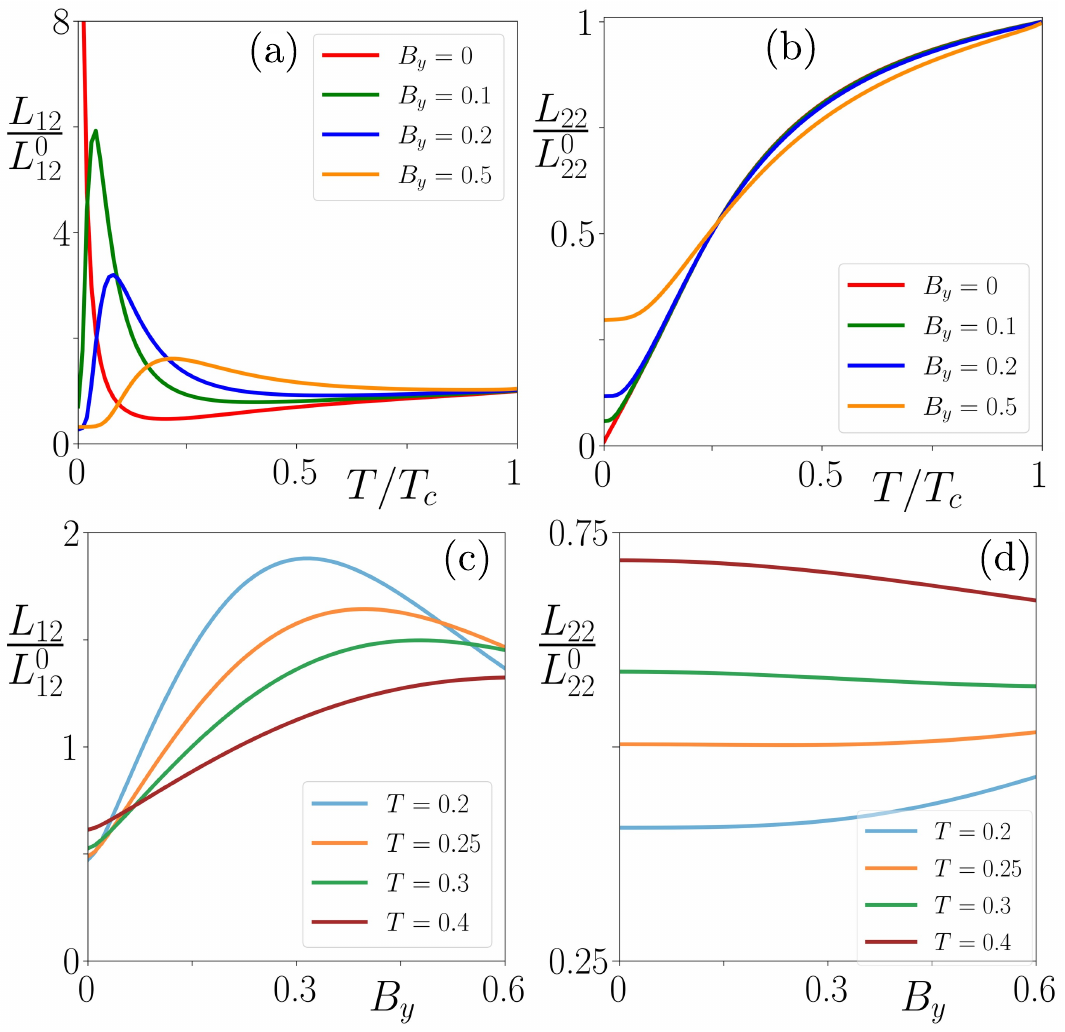}
	\caption{Normalized $L_{12}$ and $L_{22}$ are illustrated as a function of (a-b) $T/T_c$ for various values of $B_y$ and (c-d) $B_y$ choosing different values of $T/T_c$ by setting $\alpha=\pi/4$ 
	and $Z=10$ in the tunneling limit.}
	\label{Fig6_L12_L22_alpha_d_xy}
\end{figure}
We display the normalized $L_{12}$ and $L_{22}$ as a function of $T/T_c$ for various values of $B_y$ in Fig.\,\ref{Fig6_L12_L22_alpha_d_xy}(a-b), respectively. We mainly focus on the results in the tunneling limit to capture the signature of BFSs in the presence of ABSs. In Fig.\,\ref{Fig6_L12_L22_alpha_d_xy}(a), we observe a peak in $L_{12}/L_{12}^0$ close to $T/T_c=0$. It is a direct signature of ABS which was absent in case of $d_{x^2-y^2}$ pairing. Then, as we turn on the magnetic field $B_{y}$, the peak in $L_{12}/L_{12}^0$ shifts to the higher temperature regime and the peak height reduces. 
The splitting of zero-energy ABSs to finite energies is responsible for such shift. With finite $B_y$, the peak location shifts from $E=0$ to $E/\Delta_0=\pm B_y$ as discussed in Ref.~\cite{Pal2024}. Since for $B_y=0$, the ABSs are located at zero energy, the system can access those states even at $T/T_c=0$ following the Fermi-Dirac distribution function. 
However, for finite $B_y$, the ABSs are shifted to finite energies. Therefore, to access these higher energy states, system requires sufficient thermal energy. For this reason, the position of the peak shifts to higher $T/T_c$ values as the strength of $B_y$ is increased. Also, the peak height of $L_{12}$ reduces as probability of AR  due to ABSs is reduced as one increases $B_y$. Furthermore, along with the peak shift, we observe a broadening of the peak width in Fig.\,\ref{Fig6_L12_L22_alpha_d_xy}(a) for higher magnetic fields which can be primarily attributed to the generation of BFSs. Interestingly, controlling the position and localization length of ABSs in a planer Josephson junctions involving $s$-wave SCs with the help of phase bias and magnetic field has been reported in the Ref.\,\cite{Banerjee2023}. In our system, we similarly anticipate that the localization length of the ABSs can be modulated by applying an external magnetic field. Note that, in Ref.\,\cite{Banerjee2023}, since the magnetic field is applied at the junction of two conventional $s$-wave SCs, generation of BFSs is not possible and do not play any role in that case. Afterwards, we study the behaviour of the normalized $L_{22}$ as a function of $T/T_c$ in Fig.\,\ref{Fig6_L12_L22_alpha_d_xy}(b). In contrast to $L_{12}$, the presence of ABSs is not visible in $L_{22}$ and closely resembles with the results for the $d_{x^2-y^2}$ pairing [see Fig.\,\ref{Fig2.L12_L22_vs_temp}(c-d)]. Behaviour of $L_{22}$ with temperature for zero magnetic field, has been reported in Refs.~\cite{Devyatov2000,Yokoyama2005}. We investigate the effect of $B_y$ on $L_{12}$ and $L_{22}$ to understand the role of BFSs in thermoelectric properties. We show the behaviour of normalized $L_{12}$ and $L_{22}$ as a function of magnetic field in Fig.\,\ref{Fig6_L12_L22_alpha_d_xy}(c) and \ref{Fig6_L12_L22_alpha_d_xy}(d) respectively, for fixed values of $T/T_c$. In Fig.~\ref{Fig6_L12_L22_alpha_d_xy}(c), we observe a substantial rise in $L_{12}$ as we increase the magnetic field compared to Fig.\,\ref{Fig6_L12_L22_alpha_d_xy}(d) where $L_{22}$ increases at a slower rate with magnetic field at lower temperatures (for $T/T_c=0.2$). This intricate behavior can be explained by analyzing the behavior of the transmission function, $\mc{T}_e(E)$ and $\mc{T}_{th}(E)$. Note that, for $d_{xy}$ pairing, we cannot ignore the contribution arising due to the AR probability, $R_{\sg}^h(E)$, as we did in case of $d_{x^2-y^2}$ pairing [see Eq.~\eqref{Eq.T_q_Z10_dx2y2}]. This is due to the formation of ABSs which lead to the multiple ARs~\cite{Kashiwaya2000} at the interface. Therefore, utilizing the unitarity relation [see Eq.~\eqref{Eq.Unitarity}], one can express $\mc{T}_e(E)$ of Eq.~\eqref{Eqn.T_e(E)} and $\mc{T}_{th}(E)$ of Eq.~\eqref{Eqn.T_th(E)} as, 

\begin{figure}
	\includegraphics[scale=0.47]{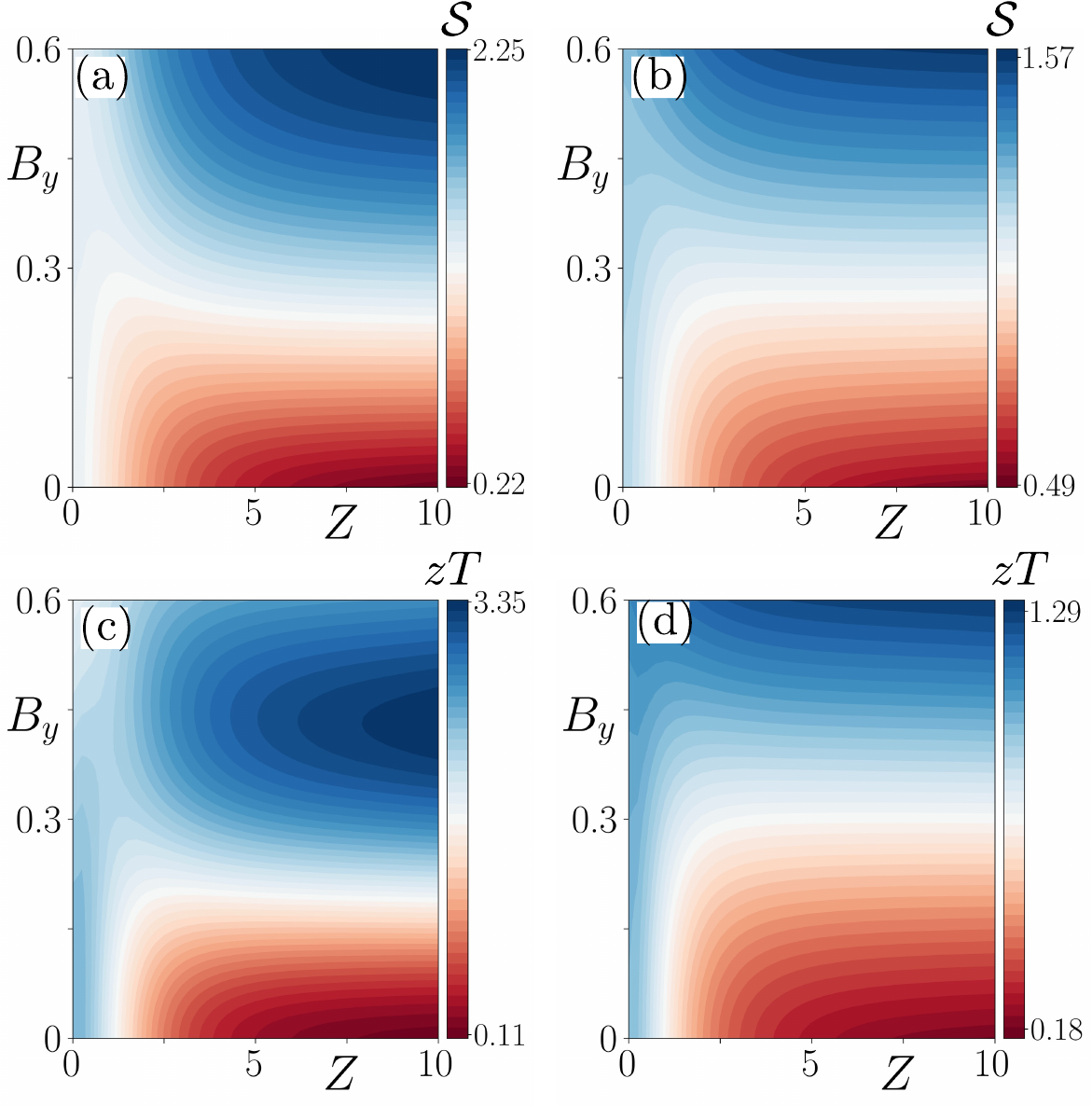}
	\caption{(a-b) Seebeck coefficient (measured in units of $k_B/e$) and (c-d) figure of merit, $zT$, are presented in the $B_y\!-Z$ plane choosing $T/T_c=0.2$ (a,c) and $T/T_c=0.4$ (b,d) considering $\alpha=\pi/4$ case.}
	\label{Fig7.S_zT_d_xy}
\end{figure}
\begin{eqnarray}
	\mc{T}_{e}(E) \!&=&\!\!\!\! \sum_{\sg}\int_{-\pi/2}^{\pi/2}\!\!\! d\theta_e \cos\theta_e [\,2\mc{R}_{\sg}^h(E) + \mc{T}_{\sg}^e(E)]\ , \label{Eqn. T_e(E)_dxy}\\
	\mc{T}_{th}(E) \!&=&\!\!\! \sum_{\sg}\int_{-\pi/2}^{\pi/2}d\theta_e \cos\theta_e [\, \mc{T}_{\sg}^e(E)]  \label{Eqn. T_th(E)_dxy}\ .
\end{eqnarray}
We note that, the behaviour of $\mc{T}_{th}(E)$ is similar for both $d_{x^2-y^2}$ and $d_{xy}$ pairing in the tunneling limit. However, interestingly, $\mc{T}_{e}(E)$ acquires an extra component, 
$2\mc{R}_{\sg}^h(E)$, due to nonzero probability of the AR from the ABSs. This leads to the peak in $L_{12}/L_{12}^0$ for $B_y=0$ [see Fig.\,\ref{Fig6_L12_L22_alpha_d_xy}(a)]. As pointed out earlier, 
the generation of BFSs for finite $B_y$ leads to the enhancement of $\mc{T}_{\sg}^e$ which further increases $\mc{T}_{e}(E)$. Due to this reason, we observe a significant rise in $L_{12}$ as we 
increase the magnetic field [see Fig.\,\ref{Fig6_L12_L22_alpha_d_xy}(c)]. Such enhancement has a major impact in Seebeck coefficient ($\mc{S}$) and figure of merit ($zT$) which we discuss in the upcoming subsection. Moreover, similar to $d_{x^2-y^2}$ pairing, the behaviour of $\mc{T}_{th}(E)$ leads to qualitatively equivalent results for the normalized $L_{12}$ and $L_{22}$ in case of $d_{xy}$ pairing in the tunneling limit [see Fig.\,\ref{Fig6_L12_L22_alpha_d_xy}(b) and Fig.\,\ref{Fig6_L12_L22_alpha_d_xy}(d)].

\subsubsection{Seebeck coefficient and figure of merit}
Now, we discuss one of the central results of this paper. Here, we present the variation of $\mc{S}$ in the $B_y-Z$ plane in Fig.\,\ref{Fig7.S_zT_d_xy}(a) and \ref{Fig7.S_zT_d_xy}(b) considering $T/T_c=0.2$ and $T/T_c=0.4$, respectively. We observe a large enhancement of $S$ as we increase the magnetic field for both regime of $T/T_c$. For a fixed $B_{y}>0.3$ such enhancement becomes larger with the increase in the barrier strength $Z$ as both $\mc{R}_{\sg}^h(E)$ and $\mc{T}_{\sg}^e$ contribute to $\mc{T}_{e}(E)$ in the tunneling limit. Large Seebeck coefficients $\sim 2.25 k_B/e$ ($\sim 200 \mu V/K$) and $\sim 1.6 k_B/e$ ($\sim 140 \mu V/K$) are observed for $T/T_c=0.2$ and $T/T_c=0.4$ respectively. Therefore, for $d_{xy}$ pairing, the generation of BFSs improves the conversion rate of heat energy into electric current which is one of the main purposes of any desirable thermoelectric device. In addition to $S$, we also study the behaviour of $zT$ in the same plane of $B_y$ and $Z$ for fixed values of $T/T_c$ and display in Fig.\,\ref{Fig7.S_zT_d_xy}(c-d). We notice that $zT$ attains a maximum value of $3.35$ for $T/T_c=0.2$ and $\sim 1.3$ for $T/T_c=0.4$. Interestingly, the maximum value of $zT$ is achieved in the tunneling limit with $B_y \approx 0.45\Delta_0$ when $T/T_c=0.2$ [see Fig.\,\ref{Fig7.S_zT_d_xy}(c)]. This is very close to the value of $Z$ and $B_y$ where $\mc{S}$ also becomes maximum. This enhancement of $zT$ relies on the maximization of the electrical conductance, $L_{11}$ too and also the thermoelectric coefficient, while minimizing the thermal conductance [see Eq.~\eqref{Eq. zT}]. From Eqs.~\eqref{Eqn. T_e(E)_dxy} and \eqref{Eqn. T_th(E)_dxy}, we observe that $\mc{T}_e(E)$ carry contributions from both ABSs and BFSs, whereas $\mc{T}_{th}(E)$ has contribution only from BFSs. Presence of ABS leads to multiple ARs~\cite{Kashiwaya2000} which enhances $\mc{T}_e(E)$ only. There is no effect of ABSs on $L_{22}$ as $\mc{R}_{\sg}^h(E)$ is absent in Eq.~\eqref{Eqn. T_th(E)_dxy}. As a consequence, this leads to an enhancement of $L_{12}$ over $L_{22}$ as $T/T_c $ is varied [see Fig.\ref{Fig6_L12_L22_alpha_d_xy}(a)-(b)]. For finite values of $B_y$, BFSs are generated in the system and existing zero energy ABSs are splitted to finite energies, $E/\Delta_0=\pm B_y$~\cite{Pal2024}. Generation of BFSs leads to an increase in $\mc{T}_{\sg}^e$ which enhances both $\mc{T}_{e}$ and $\mc{T}_{th}$ [See Eq. \eqref{Eqn. T_e(E)_dxy}-\eqref{Eqn. T_th(E)_dxy}]. However, due to the generation of BFSs together with finite energy ABSs, $L_{12}$ increases more rapidly compared to $L_{22}$ [see Fig.\ref{Fig6_L12_L22_alpha_d_xy}(c)-(d)]. Due to these reasons, presence of BFSs leads to large values of $zT$ for $d_{xy}$ pairing SC. Therefore, after investigating both $\mc{S}$ and $zT$, we find our hybrid junction, hosting BFSs as zero energy excitations, can be a potential platform for thermoelectric applications.

\begin{figure}[]
	\includegraphics[scale=0.48]{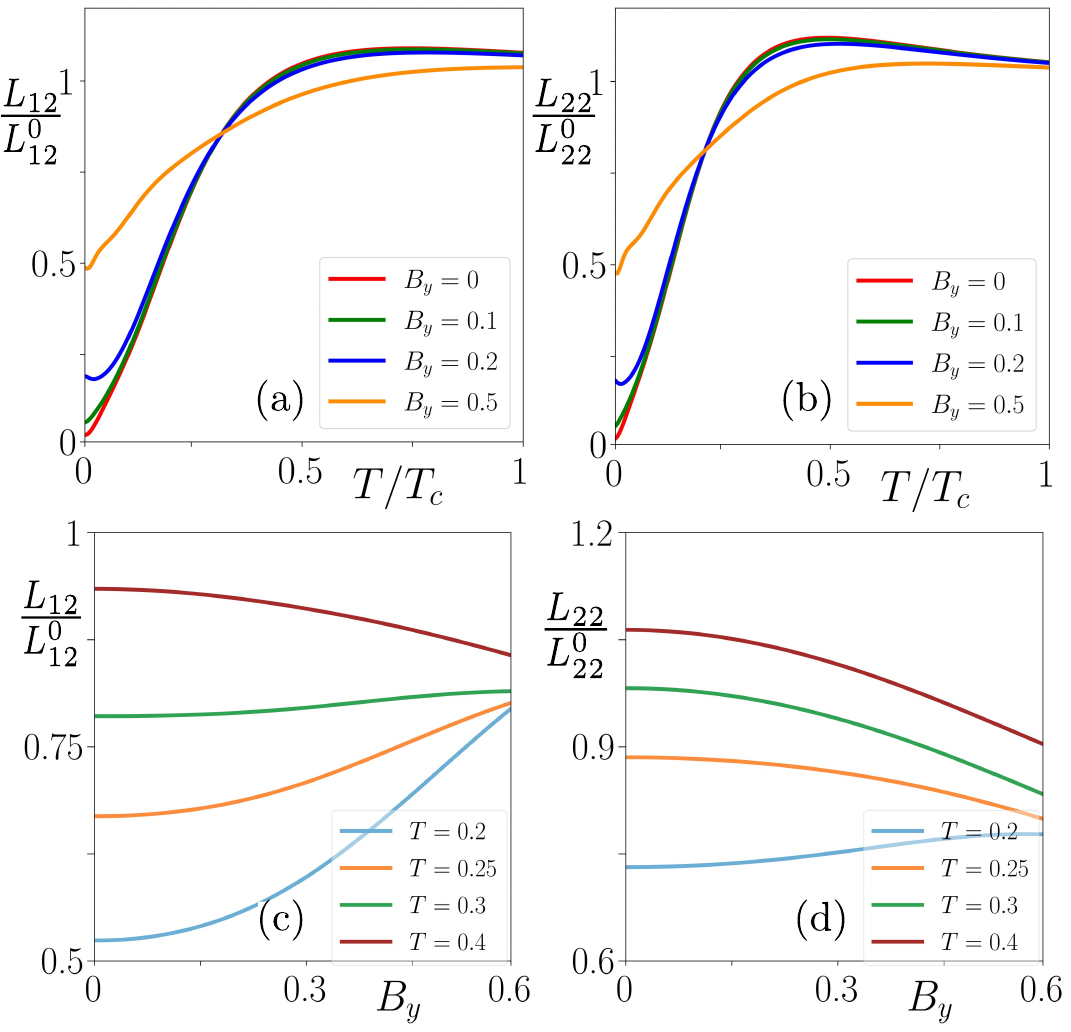}
	\caption{In panels (a-b), normalized $L_{12}$ and $L_{22}$ are depicted as a function of
		$T/T_c$ for various strengths of $B_y$ and in panels (c-d), the same have been displayed as a 
		function of $B_y$ choosing different values of $T/T_c$ in the tunneling limit. We consider the 
		system size as $(N_x,N_y)=(100a,250a)$ with $a$ ($=1)$ being the lattice spacing along both  
		$x$ and $y$ directions contemplating $\alpha=0$ and $Z=10$.}
	\label{Fig8.Lattice_results}
\end{figure}
\subsection{Lattice model}\label{Subsec:III_c}
With the investigation of the signatures and role of BFSs in the thermoelectricity using a continuum model, in this subsection, we examine the thermoelectric response based on a lattice model simulation. To incorporate the lattice model, we first discretize the continuum model in a square lattice to obtain the lattice regularized tight-binding Hamiltonian as,
\begin{eqnarray}
\mc{H}_{{\rm{lat}}}(\mathbf{k})\!&=&\![4t-\mu-2t(\cos k_x + \cos k_y) ]  \tau_z  \sg_0 -B_y \tau_0  \sg_y  \non \\ && + \Del_{{\rm{lat}}}(\mathbf{k},\alpha) \tau_x \sg_0\ ,
\label{Eq.lattice_model}
\end{eqnarray}
where, $\Del_{{\rm{lat}}}(\mathbf{k},\alpha=0)=2\Del_0(\cos k_y - \cos k_x)$ and $\Del_{{\rm{lat}}}(\mathbf{k},\alpha=\pi/4)= 2\Delta_0\sin k_x\sin k_y$~\cite{Setty2020PRB,Yang1998,Pal2024}. We refer to Appendix~\ref{Append_B} for more details to obtain the tight-binding model and transmission functions. Using the Python package KWANT~\cite{Groth2014}, we compute the transmission functions, $\mc{T}_e(E)$ and $\mc{T}_{th}(E)$ [see Eqs.~\eqref{Eqn.T_e(E)} and \eqref{Eqn.T_th(E)})] for this lattice model. Then, we numerically integrate the transmission functions following Eqs.~\eqref{Eqn. L12} and \eqref{Eqn. L22} to obtain the thermoelectric coefficient and thermal conductance. In Fig.~\ref{Fig8.Lattice_results}(a) and (b), we present our results for the normalized $L_{12}$ and $L_{22}$ as a function of $T/T_c$ choosing various strengths of $B_y$ 
in the tunneling limit for $\alpha=0$. We also compute the normalized $L_{12}$ and $L_{22}$ as a function of $B_y$ in the tunneling limit for various values of $T/T_c$ and depict them in Fig.~\ref{Fig8.Lattice_results}(c-d). Comparing Fig.~\ref{Fig8.Lattice_results} with Fig.~\ref{Fig2.L12_L22_vs_temp}(b-d) and Fig.~\ref{Fig3.L12_L22_vs_By}(b-d), we find excellent 
agreement between the continuum and lattice model. 

\section{Summary and Conclusions}\label{Sec:IV_Summary}
To summarize, in this article, we have systematically investigated the thermoelectric response of BFSs, within the BTK formalism~\cite{Blonder1982}, considering a $d$-wave SC hybrid junction under the application of a temperature gradient across the system. In our set up, BFSs are generated by applying an in-plane magnetic field in an unconventional $d$-wave SC. An external gate voltage at the junction can tune the junction transparency. We compute the thermoelectric coefficient, thermal conductance, Seebeck coefficient, figure of merit, and examine the validation of WF law for various sets of parameter values. Additionally, changing the orientation angle of the $d$-wave vector, $\alpha$, from zero to $\pi/4$ leads to the formation of ABSs in the interface giving rise to a peak in $L_{12}$. Interestingly, for $\alpha=\pi/4$, we observe a significant enhancement in the thermopower due to the generation of BFSs with maximum value of $S\sim 2.25k_B/e \sim 200 \,\mu V/K$ at $T/T_c=0.2$ which is significantly larger than the thermopower in metals (in order of $\mu V/K$)~\cite{Kolenda2016}. 
This enhancement of the thermopower implies a large rate of conversion from the heat energy to the electric current. 
Therefore, one can achieve higher thermoelectricity using a $d$-wave SC in presence of a magnetic field hosting BFSs compared to the case when magnetic field is absent in the system \cite{Devyatov2000,Yokoyama2005}. Concomitantly, we observe high value of $zT\sim 3.5$ which makes our proposed heterostructure a potential platform for thermoelectric applications. Furthermore, we strengthen our continuum model results by computing the thermoelectric responses based on a lattice regularized tight binding model using the Python package KWANT~\cite{Groth2014}. 

Note that, our proposed model, based on the $d$-wave SC, serves as the minimal model exhibiting BFSs. In reality, $d$-wave SC hosts multiband pairing with both inter- and intra-band orbital pairing amplitudes originating from strong electron-electron interactions~\cite{Setty2020NatComm}. However, our purpose is to identify the basic signatures of BFSs via thermoelectric response in a single band model which can be achieved without incorporating the multiband pairing 
potentials~\cite{Yuan2018,Banerjee2022,Pal2024}. Therefore, our study provides the primary thermoelectric signatures of BFSs and the possibility of its usage as an efficient thermoelectric device component. 

\subsection*{Acknowledgments}
A.P. acknowledges Pritam Chatterjee for stimulating discussions. A.P. and A.S. acknowledge SAMKHYA: High-Performance Computing facility provided by Institute of Physics, Bhubaneswar and the Workstation provided by Institute of Physics, Bhubaneswar from DAE APEX Project for numerical computations. P.\,D. thanks Mohit Randeria for helpful discussion at the initial stage of this work. 
P.\,D. acknowledges Department of Space (DoS), India for all support at PRL and Anusandhan National Research Foundation (ANRF) erstwhile Science and Engineering Research Board (SERB), Department of Science and Technology (DST), India for the financial support through Start-up Research Grant (File no.\,SRG/2022/001121).
\begin{appendix}

\section{Scattering matrix formalism}\label{Appendix_scattering_Matrix}
In this appendix, we provide our basic theoretical framework applied for the continuum model. 
In a typical normal metal-SC junction, depending on the energy of incident electron, the electron can scatter from the interface via the following quantum mechanical processes, (i) normal reflection (NR): reflection as a normal electron, (ii) Andreev reflection (AR): reflection as a hole through formation of a Cooper pair that jumps into the SC, and (iii) transmission as an electron-like and hole-like quasiparticles. Interplay of these scattering processes are the root cause of various physical phenomena. We consider the transport along the $x$-direction and electrons are incident from the normal metal side ($x<0$). In order to investigate further, first we solve the BdG Hamiltonian in 
Eq.~\eqref{Eq:Model_Ham} to obtain the single-particle energy eigenstates for a given excitation energy, $E$. In the normal metal side ($x<0$), the energy eigenstates are given by, 

\vspace{-0.5cm}
\begin{widetext}
\begin{equation*}
	\psi^{{\rm{inc}}}_{e,\UP}(x) \!=\! \frac{1}{\sqrt{2}}\begin{bmatrix}
		~ 1 ~\\ i\\ 0 \\ 0 
	\end{bmatrix} e^{ik_e\cos\theta_e x}\ ,~~
	\psi^{{\rm{inc}}}_{e,\DN}(x) \!=\! \frac{1}{\sqrt{2}}\begin{bmatrix}
		~ i ~\\ 1\\ 0 \\ 0 
	\end{bmatrix} e^{ik_e\cos\theta_e x}\ ,
\end{equation*}


\begin{equation*}
	\psi^{{\rm{ref}}}_{e,\UP}(x)\! =\!\frac{1}{\sqrt{2}} \begin{bmatrix}
		~ 1 ~\\ i\\ 0 \\ 0 
	\end{bmatrix}\! e^{-ik_e\cos\theta_e x} ~~,~~
	\psi^{{\rm{ref}}}_{e,\DN}(x)\! =\!\frac{1}{\sqrt{2}} \begin{bmatrix}
		~ i ~\\ 1\\ 0 \\ 0 
	\end{bmatrix}\! e^{-ik_e\cos\theta_e x}~~,
\end{equation*}

\begin{equation*}
	\psi^{{\rm{ref}}}_{h,\DN}(x)\! =\! \frac{1}{\sqrt{2}}\begin{bmatrix}
		~ 0 ~\\ 0\\ 1 \\ i 
	\end{bmatrix} e^{ik_h\cos\theta_h x}~~,~~
	\psi^{{\rm{ref}}}_{h,\UP}(x) \!=\! \frac{1}{\sqrt{2}}\begin{bmatrix}
		~ 0 ~\\ 0\\ i \\ 1 
	\end{bmatrix} e^{ik_h\cos\theta_h x} ~~, 
\end{equation*} 
\end{widetext}
where, $\psi^{{\rm{inc(ref)}}}_{\alpha,\sg}(x)$ represents the incident (reflected) state with spin $\sg=\{\UP,\DN\}$, and particle type, $\alpha=\{e,h\}$. The momentum of the incident electron and reflected hole is given as, $k_{e(h)}=\sqrt{2m(E+(-)\mu)}$ respectively. The angle $\theta_e$ and $\theta_h$ correspond to the electron incident and hole reflection angle, respectively. Using the conservation of the parallel component of the momentum i.e $k_e\sin\theta_e=k_h\sin\theta_h$, we can express the reflection angle of hole, $\theta_h$, in terms of the incident angle of electron, $\theta_e$, as 
$\displaystyle{\theta_h=\sin^{-1}\left(\sqrt{\frac{\mu + E}{\mu-E}} \sin\theta_e\right)}$. 

Similarly, in the SC side ($x>0$), the eigenstates are obtained as,
\begin{equation*}
	\psi^{{\rm{trans}}}_{eL,\UP}(x) = \frac{1}{\sqrt{2}}\begin{bmatrix}
		~ u_\UP(\theta_+) ~\\ i\,u_\UP(\theta_+)\\i\, v_\UP(\theta_{+})e^{-i\phi_{+}} \\ v_\UP(\theta_{+})e^{-i\phi_{+}} 
	\end{bmatrix} e^{ik_{eL,\UP}x\cos\theta_{eL,\UP} }\ ,
\end{equation*}
%
\begin{equation*}
	~~\psi^{{\rm{trans}}}_{eL,\DN}(x) = \frac{1}{\sqrt{2}}\begin{bmatrix}
		~i\, u_\DN(\theta_+)~\\u_\DN(\theta_+) \\ v_\DN(\theta_+)e^{-i\phi_+}\\ i\,v_\DN(\theta_+)e^{-i\phi_+} 
	\end{bmatrix} e^{ik_{eL,\DN}x\cos\theta_{eL,\DN} }\ ,
\end{equation*}
\begin{equation*}
	\psi^{{\rm{trans}}}_{hL,\UP}(x) =\frac{1}{\sqrt{2}} \begin{bmatrix}
		~ v_\UP(\theta_-) \\ i\,v_\UP(\theta_-)\\i\, u_\UP(\theta_{-})e^{-i\phi_{-}} \\  u_\UP(\theta_{-})e^{-i\phi_{-}} 
	\end{bmatrix} e^{-ik_{hL,\UP}x\cos\theta_{hL,\UP} }\ ,
\end{equation*}
\begin{equation*}
	~~\psi^{{\rm{trans}}}_{hL,\DN}(x) = \frac{1}{\sqrt{2}}\begin{bmatrix}
		~ i\,v_\DN(\theta_-)\\v_\DN(\theta_-) \\  u_\DN(\theta_-)e^{-i\phi_-} \\ i\,u_\DN(\theta_-)e^{-i\phi_-} 
	\end{bmatrix} e^{-ik_{hL,\DN}x\cos\theta_{hL,\DN} }\ ,
\end{equation*}
where, $\psi^{{\rm{trans}}}_{eL(hL),\sg}$ represents the transmitted electron-like (hole-like) quasi-particle state with spin $\sg$. The momenta of the transmitted electron and hole like states with spin $\sg$ are given by, $k_{eL(hL),\sg}\!=\!\sqrt{2m \left(\mu +(-)\sqrt{(E-\sg B)^2-|\Delta(\theta_{+(-)})|^2}\right)}$. The superconducting coherence factors are given as follow:
\begin{subequations}
	\begin{align}
		\dis{u_\sg(\theta_\pm)=\frac{1}{\sqrt{2}} \left[ 1 + \sqrt{\frac{(E-\sg B)^2 - |\Delta(\theta_\pm)|^2}{(E-\sg B)^2}}\right]^{1/2}}\ , \non  \\
		\dis{v_\sg(\theta_\pm)=\frac{1}{\sqrt{2}} \left[ 1 - \sqrt{\frac{(E-\sg B)^2 - |\Delta(\theta_\pm)|^2}{(E-\sg B)^2}}\right]^{1/2}}\ , \non 
	\end{align}
\end{subequations}
where the pair potential is given as,
\begin{equation*}
	\Delta(\theta_\pm)=\Delta_0\cos(2\theta\mp2\alpha)\ ,
\end{equation*}
satisfying the relation $\dis{e^{i\phi_\pm}=e^{i\phi} \frac{\Delta(\theta_\pm)}{|\Delta(\theta\pm)|}}$.
The transmission angles for the electron and hole like quasiparticles inside the SC can be obtained by employing the conservation of the $y$-component of wave-vector \ie 
$k_{eL,\sg}\sin\theta_{eL,\sg}=k_{e}\sin\theta_e\ $ and 
$k_{hL,\sg}\sin\theta_{hL,\sg}=k_{e}\sin\theta_e\ $, and obtained as,
\begin{subequations}
	\begin{align}
		\dis{\theta_{eL,\sg}=\sin^{-1}\left(\sqrt{\frac{\mu + E}{\mu + \sqrt{(E-\sg B)^2 -|\Delta(\theta_+)|^2}}} \sin\theta_e \right)}\ , \non  \\
		\dis{\theta_{hL,\sg}=\sin^{-1}\left(\sqrt{\frac{\mu + E}{\mu - \sqrt{(E-\sg B)^2 -|\Delta(\theta_-)|^2}}} \sin\theta_e \right)}\ \non \ .
	\end{align}
\end{subequations}

Therefore, the total wavefunction in both normal metal side ($x<0$) and SC side $(x>0)$ can be written as, 
\begin{eqnarray}
	\Psi^N_\sg(x) \!\!&=&\!\!\! \psi^{{\rm{inc}}}_{e,\sg} + r^{ee}_{\sg,\sg}\, \psi^{{\rm{ref}}}_{e,\sg} \!+\! r^{ee}_{\bar{\sg},\sg}\, \psi^{{\rm{ref}}}_{e,\bar{\sg}} \!+\! r^{eh}_{\bar{\sg},\sg}\, \psi^{{\rm{ref}}}_{h,\bar{\sg}} \!+\!
	r^{eh}_{\sg,\sg}\, \psi^{{\rm{ref}}}_{h,\sg} \ ,\non \\ \\
	\Psi^{SC}_\sg (x) \!\! &=&\!\!\! t^{ee}_{\sg,\sg}\,\psi^{{\rm{trans}}}_{eL,\sg} +
	t^{ee}_{\bar{\sg},\sg}\,\psi^{{\rm{trans}}}_{eL,\bar{\sg}} + t^{eh}_{\sg,\sg}\,\psi^{{\rm{trans}}}_{hL,\sg}, +
	t^{eh}_{\bar{\sg},\sg}\,\psi^{{\rm{trans}}}_{hL,\bar{\sg}}\ , \non \\ 
\end{eqnarray}
where, $r^{ee}_{\bar{\sg},\sg},r^{eh}_{\bar{\sg},\sg},t^{ee}_{\bar{\sg},\sg},t^{eh}_{\bar{\sg},\sg}$ designate the scattering amplitudes for the NR, AR, electron-like, and hole-like quasiparticle transmission, respectively, for an incident electron with spin $\sg$ to a scattered state with spin 
$\bar{\sg}$. We obtain the solution of these equations employing the following boundary conditions, 
\begin{eqnarray}
	\Psi^N_\sg(x=0) &=& \Psi^{SC}_\sg(x=0) \label{Eqn.NS}\ , \nonumber \\ 
	\frac{d}{dx}\Psi^{SC}_\sg(x) \vline_{\,x=0^+}-\frac{d}{dx}\Psi^{N}_\sg(x) \vline_{\,x=0^-}  &=& Z\Psi^N_\sg(x=0)\ , \nonumber \\ 
	\label{Eqn.NS_deriv}
\end{eqnarray}
Here, $Z=\frac{2mV_B}{k_f}$ is the dimensionless barrier strength which may be externally tuned using a gate voltage, applied at the junction. In our study, we coin $Z=0$ and $Z=10$ as ballistic and tunneling limit, respectively.

\vspace{+0.2cm}
\section{Contribution of sub-gap and above-gap states} \label{Append_A}
In this appendix, we discuss the contribution arising due to sub-gap states \ie QPs with energy $E<\Del_0$ and above gap states i.e., QPs with energy $E>\Del_0$ to $L_{12}$ and $L_{22}$ by dividing the integration limit of Eq.~\eqref{Eqn. L12} and Eq.~\eqref{Eqn. L22} into two parts as, 
\vspace{+0.5cm}
\begin{widetext}
\begin{eqnarray*}
L_{12}&=& \frac{1}{T}\int_{0}^{\infty} dE \,\mc{T}_e(E) E \left(-\frac{\partial f (E,T)}{\partial E}\right)   
\ , \\
&=& \frac{1}{T}\int_{0}^{\Del_0} dE \,\mc{T}_e(E) E \left(-\frac{\partial f (E,T)}{\partial E}\right) + \frac{1}{T}\int_{\Del_0}^{\infty} dE \,\mc{T}_e(E) E \left(-\frac{\partial f (E,T)}{\partial E}\right) \ ,  \\
\implies L_{12}&=& L_{12}^< + L_{12}^>\ ,  \\
L_{22}&=& \frac{1}{T} \int_{0}^{\infty} dE \,\mc{T}_{th}(E) E^2 \left(-\frac{\partial f (E,T)}{\partial E}\right)\ ,  \\
&=&\frac{1}{T} \int_{0}^{\Delta_0} dE \,\mc{T}_{th}(E) E^2 \left(-\frac{\partial f (E,T)}{\partial E}\right)  +
\frac{1}{T} \int_{\Del_0}^{\infty} dE \,\mc{T}_{th}(E) E^2 \left(-\frac{\partial f (E,T)}{\partial E}\right)\ , \\
\implies L_{22} &=& L_{22}^< + L_{22}^>\ .
\end{eqnarray*}
\end{widetext}

\begin{figure}[]
	\includegraphics[scale=0.42]{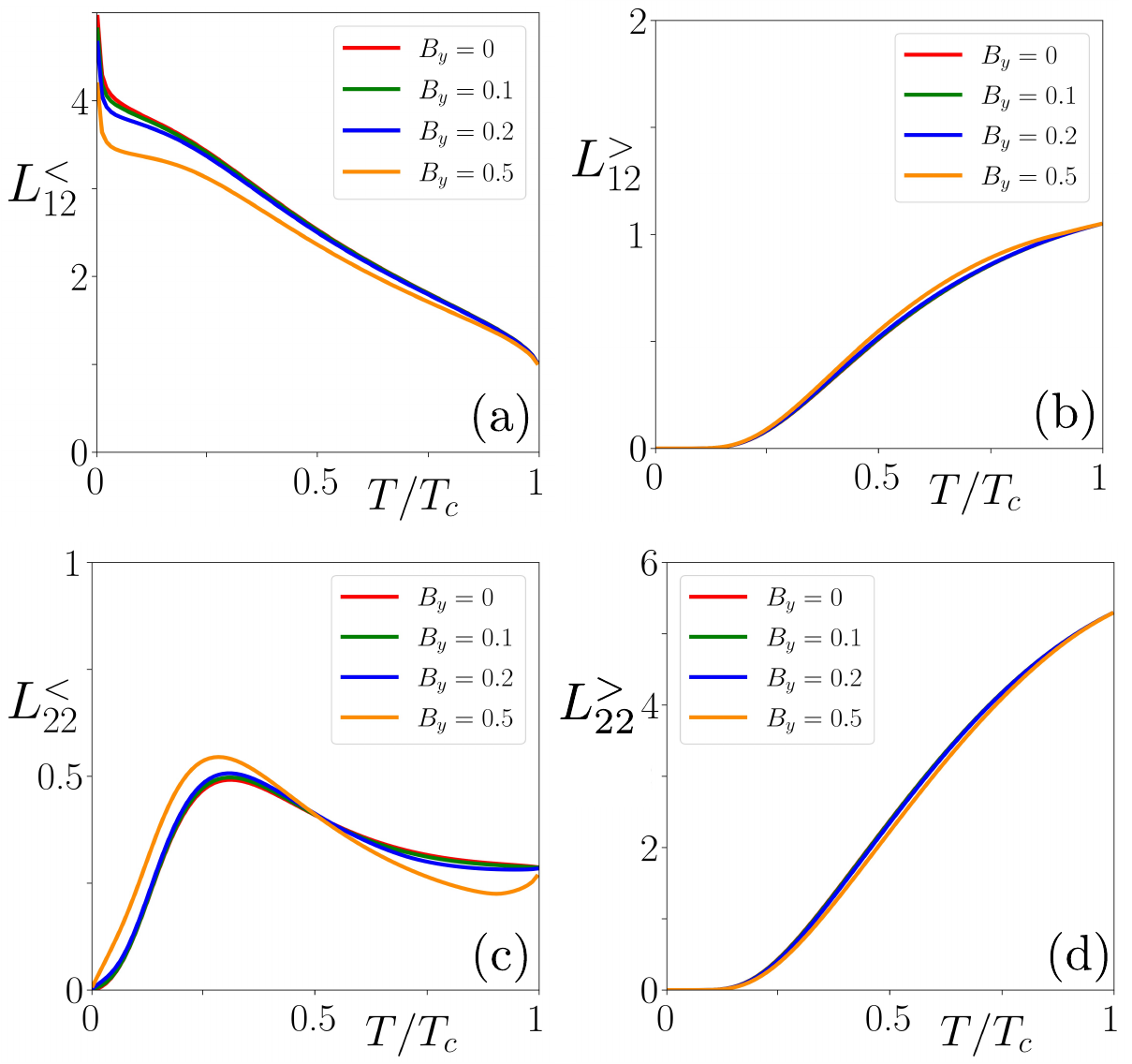}
	\caption{In panels (a), (b) we illustrate the behaviour of $L_{12}^<$ and $L_{12}^>$ respectively, as a function of $T/T_c$ for various values of $B_y$. In panels (c), (d) we showcase 
the variation of $L_{22}^<$ and $L_{22}^>$ respectively, as a function of $T/T_c$ choosing the same values of $B_{y}$. We set $\alpha=0$ and $Z=0$ in all the panels. Rest of the model parameters remain the same as mentioned in the main text.}
	\label{Fig.A1}
\end{figure}

Since BFSs are the zero-energy excitations in the system, their contribution can be captured in $L_{12}^<$ and $L_{22}^<$. Whereas, contributions arising due to states with energy $E>\Del_0$ will be reflected in $L_{12}^>$ and $L_{22}^>$. We depict the behaviour of $L_{12}^<$, $L_{12}^>$,$L_{22}^<$, and $L_{22}^>$ as a function of $T/T_c$ in Fig.~\ref{Fig.A1}(a-d) for $Z=0$ (transparent limit) and $\alpha=0$. We observe that for low enough temperatures (below $T/T_c \sim 0.25$), only $L_{12}^<$ and $L_{22}^<$ exhibit finite values and significantly contribute to $L_{12}$ and $L_{22}$. Note that, the behavior of $L_{12}^<$ and $L_{22}^<$ as a function of $T/T_c$ qualitatively dictates the behavior of $L_{12}$ [see Fig.\ref{Fig2.L12_L22_vs_temp}(a)] and $L_{22}$ [see Fig.\ref{Fig2.L12_L22_vs_temp}(c)] respectively, within that temperature regime. At $T\rightarrow0$, the reduction in $L_{12}^<$ and enhancement in $L_{22}^<$ with respect to $B_{y}$ occur due to the generation of BFSs. Due to this reason, to understand the signatures of only BFSs, 
we compute $\mc{S}$ and $zT$, in the main text, at lower values of $T/T_c$. On the contrary, the contribution of $L_{12}^>$ and $L_{22}^>$ become nonzero above $T/T_c>0.25$. This behaviour has the roots in Fermi-Dirac distribution functions and its dependency in $L_{12}$ [see Eq.~\eqref{Eqn. L12}] and $L_{22}$ [see Eq.~\eqref{Eqn. L22}]. Therefore, only for sufficiently large $T/T_c$, the occupation of the states with energy $E>\Del_0$ becomes more probable and contribute to the thermoelectric properties of the system. This can also be the reason behind the existence of cross-over temperature, $T_r$, above which the QPs with $E>\Del_0$ contribute substantially and screen the signatures of BFSs. In a similar fashion, one can also analyse $L_{12}^<$ and $L_{22}^<$
in the tunneling limit (\eg $Z=10$). 

\section{Lattice model calculations} \label{Append_B}

In this appendix, we discuss the details of our calculation for the lattice regularized version of the continuum model [see Eq.\eqref{Eq:Model_Ham} in the main text]. First, we obtain the tight-binding Hamiltonian in a square lattice with the lattice constant $a$ (=1) by replacing $k_i^2\sim 2(1-\cos k_i)$ and $k_i\sim \sin k_i$ ($i=x,y$) as, 
\begin{eqnarray}
	\mc{H}_{{\rm{lat}}}(\mathbf{k})\!&=&\![4t-\mu-2t(\cos k_x + \cos k_y) ]  \tau_z  \sg_0 -B_y \tau_0  \sg_y  \non \\ && + \Del_{{\rm{lat}}}(\mathbf{k},\alpha) \tau_x \sg_0\ , \non 
\end{eqnarray}
where, $\Del_{{\rm{lat}}}(\mathbf{k},\alpha=0)=2\Del_0(\cos k_y - \cos k_x)$ and $\Del_{{\rm{lat}}}(\mathbf{k},\alpha=\pi/4)= 2\Delta_0\sin k_x\sin k_y$. Then, we choose a rectangular geometry with dimension $N_x$ and $N_y$ along $x$- and $y$-directions. For $x\geq N_x/2$, the system is \scing ($\Delta_0\neq 0$) and hopping amplitudes and onsite potentials are governed by the above Hamiltonian with $B_{y}\neq 0$. On the other hand, for $x\le N_x/2$, the system is in the normal state i.e., $\Delta_0=0$ and $B_y=0$. We attach a metallic lead at $x=0$ and a \scing lead at $x=N_x$ to obtain the transport properties of the system along $x$-direction. We model the \scing lead by the same Hamiltonian used to describe our system. Afterwards, from Eqs.~\eqref{Eqn. L12} and \eqref{Eqn. L22}, it is evident that $L_{12}$ and $L_{22}$ incorporate the microscopic details of the model through the transmission functions $\mc{T}_e(E)$ and $\mc{T}_{th}(E)$ [see Eqs.~\eqref{Eqn.T_e(E)} and \eqref{Eqn.T_th(E)}]. We next employ the 
BTK formalism~\cite{Blonder1982,Groth2014} to compute these transmission functions as, 
\begin{eqnarray*}
	\mc{T}_e(E)= N_m(E) - R_{ee}(E) + R_{he}(E)\ , \\
	\mc{T}_{th}(E)= N_m(E) - R_{ee}(E) - R_{he}(E)\ , \\	
\end{eqnarray*}
where, $R_{ee}=Tr[r_{ee}^\dagger r_{ee}]$, $R_{he}=Tr[r_{he}^\dagger r_{he}]$; $r_{ee}$ and $r_{he}$ are matrices with dimension $2N_m\times 2N_m$ where $N_m$ is the number of occupied transverse channels in the left lead at energy, $E$. Using the Python package, KWANT~\cite{Groth2014}, we calculate the $r_{ee}$ and $r_{he}$ matrices to find $\mc{T}_e(E)$ and $\mc{T}_{th}(E)$ as a function of energy $E$. We then perform the numerical integration to compute 
$L_{12}$ and $L_{22}$ using Eqs.~\eqref{Eqn. L12} and \eqref{Eqn. L22}. 
\end{appendix}
\twocolumngrid
\bibliographystyle{apsrev4-2mod}
\bibliography{bibfile}{}

\end{document}